\documentclass[twocolumn]{aastex62}
\usepackage{epstopdf}
\usepackage{wrapfig}
\usepackage{lipsum} 
\usepackage[utf8]{inputenc}
\usepackage{amsmath}
\usepackage{natbib}
\bibpunct{(}{)}{;}{a}{}{,}

\newcommand\numberthis{\addtocounter{equation}{1}\tag{\theequation}}

\usepackage{color}
\definecolor{darkgreen}{RGB}{0,142,128}
\definecolor{darkblue}{RGB}{0,100,170}
\usepackage{hyperref}
\hypersetup{colorlinks,citecolor=blue,linkcolor=blue}

\graphicspath{{./}{figures/}}

\received{September 15, 2019}
\revised{October 17, 2019}
\accepted{October 17, 2019 in ApJS}

\shorttitle{Alfvén wave dynamics}
\shortauthors{Réville et al.}
\begin{document}

\title{The role of Alfvén wave dynamics on the large scale properties of the solar wind: comparing an MHD simulation with PSP E1 data}

\author[0000-0002-2916-3837]{Victor Réville}
\affiliation{UCLA Earth Planetary and Space Sciences Department, LA, CA, USA}
\affiliation{IRAP, Universit\'e Toulouse III - Paul Sabatier,
CNRS, CNES, Toulouse, France}

\author[0000-0002-2381-3106]{Marco Velli}
\affil{UCLA Earth Planetary and Space Sciences Department, LA, CA, USA}

\author[0000-0002-4440-7166]{Olga Panasenco}
\affil{Advanced Heliophysics, Pasadena, CA 91106, USA}

\author[0000-0003-2880-6084]{Anna Tenerani}
\affil{University of Texas, Austin, TX, USA}

\author[0000-0002-2582-7085]{Chen Shi}
\affiliation{UCLA Earth Planetary and Space Sciences Department, LA, CA, USA}

\author[0000-0002-6145-436X]{Samuel T. Badman}
\affil{Physics Department, University of California, Berkeley, CA 94720-7300, USA}
\affil{Space Sciences Laboratory, University of California, Berkeley, CA 94720-7450, USA}

\author[0000-0002-1989-3596]{Stuart D. Bale}
\affil{Physics Department, University of California, Berkeley, CA 94720-7300, USA}
\affil{Space Sciences Laboratory, University of California, Berkeley, CA 94720-7450, USA}
\affil{The Blackett Laboratory, Imperial College London, London, SW7 2AZ, UK}
\affil{School of Physics and Astronomy, Queen Mary University of London, London E1 4NS, UK}

\author[0000-0002-7077-930X]{J. C. Kasper}
\affil{Climate and Space Sciences and Engineering, University of Michigan, Ann Arbor, MI 48109, USA}
\affil{Smithsonian Astrophysical Observatory, Cambridge, MA 02138, USA}

\author[0000-0002-7728-0085]{Michael L. Stevens}
\affil{Smithsonian Astrophysical Observatory, Cambridge, MA 02138, USA}

\author[0000-0001-6095-2490]{Kelly E. Korreck}
\affil{Smithsonian Astrophysical Observatory, Cambridge, MA 02138, USA}

\author[0000-0002-0675-7907]{J. W. Bonnell}
\affil{Space Sciences Laboratory, University of California, Berkeley, CA 94720-7450, USA}

\author[0000-0002-3520-4041]{Anthony W. Case}
\affil{Smithsonian Astrophysical Observatory, Cambridge, MA 02138, USA}	
\author[0000-0002-4401-0943]{Thierry {Dudok de Wit}}
\affil{LPC2E, CNRS and University of Orl\'eans, Orl\'eans, France}

\author[0000-0003-0420-3633]{Keith Goetz}
\affiliation{School of Physics and Astronomy, University of Minnesota, Minneapolis, MN 55455, USA}

\author[0000-0002-6938-0166]{Peter R. Harvey}
\affil{Space Sciences Laboratory, University of California, Berkeley, CA 94720-7450, USA}

\author[0000-0001-5030-6030]{Davin E. Larson}
\affil{Space Sciences Laboratory, University of California, Berkeley, CA 94720-7450, USA}

\author[0000-0002-0396-0547]{Roberto Livi}
\affil{Space Sciences Laboratory, University of California, Berkeley, CA 94720-7450, USA}

\author[0000-0003-1191-1558]{David M. Malaspina}
\affil{Laboratory for Atmospheric and Space Physics, University of Colorado, Boulder, CO 80303, USA}

\author[0000-0003-3112-4201]{Robert J. MacDowall}
\affil{Solar System Exploration Division, NASA/Goddard Space Flight Center, Greenbelt, MD, 20771}

\author[0000-0002-1573-7457]{Marc Pulupa}
\affil{Space Sciences Laboratory, University of California, Berkeley, CA 94720-7450, USA}

\author[0000-0002-7287-5098]{Phyllis L. Whittlesey}
\affil{Space Sciences Laboratory, University of California, Berkeley, CA 94720-7450, USA}

\begin{abstract}
During Parker Solar Probe's first orbit, the solar wind plasma has been observed in situ closer than ever before, the perihelion on November 6th 2018 
revealing a flow that is constantly permeated by large amplitude Alfvénic fluctuations. These include radial magnetic field reversals, or switchbacks, that seem to be a persistent feature of the young solar wind. The measurements also reveal a very strong, unexpected, azimuthal velocity component. In this work, we numerically model the solar corona during this first encounter, solving the MHD equations and accounting for Alfvén wave transport and dissipation. We find that the large scale plasma parameters are well reproduced, allowing the computation of the solar wind sources at Probe with confidence. We try to understand the dynamical nature of the solar wind to explain both the amplitude of the observed radial magnetic field and of the azimuthal velocities.
\end{abstract}


\section{Introduction} 
\label{intro}

Parker Solar Probe \citep[PSP hereafter,][]{FoxPSP2016} traversed its first perihelion on November 6th 2018. After a Venus gravity assist, it reached a distance of $35.7 R_{\odot}$ from the Sun, closer by almost a factor two than the minimum distance reached by the previous record holders: the Helios probes. The orbit naturally gives the spacecraft high angular velocities, so that PSP was in co-rotation and super rotation with the Sun for significant time intervals at closest approach. Its instruments suites are composed of an electromagnetic field analyzer FIELDS \citep{Bale2016}, a plasma and particle instrument SWEAP \citep[Solar Wind Electrons Alphas and Protons][]{Kasper2016}, the Wide-Field Imager for Solar PRobe
Plus \citep[WISPR][]{Vourlidas2016}, and high energy particle instruments IS$\odot$IS \citep[Integrated Science Investigation of the Sun,][]{McComas2016}.

Measurements of the first perihelion have unraveled the "youngest" solar wind observed so far, yielding surprising features. First, large scale perturbations, with an almost full 180 degree rotation of the magnetic field, are observed over a large range of frequencies. Although "switchbacks" have already been measured and discussed in the past \citep{Balogh1999,Neugebauer2013}, they are a constant feature of the fields and plasma measurements in the first encounter data. The correlation between velocities and magnetic field perturbations is consistent with Alfvén waves with a constant total magnetic field magnitude and small relative density fluctuations. They are however, almost by definition, non-transverse and their properties may be different from usual purely perpendicular Alfvén waves. Moreover, measurements made at 1 au have shown that most of these structures have disappeared before reaching Earth orbit \citep{Panasenco2020ApJS}. Hence, if switchbacks are regular features in the young solar wind (as seems to be the case also in encounters 2 and 3), they may contain precious new information about the origin of the solar wind.

The second major surprise is what seems to be a very extended co-rotation of the solar wind \citet{Kasper2019Nature}, with rotational velocities up to some 50 to 70 km/s at perihelion. These measurements were obtained by the Solar Probe Cup (SPC), the Faraday cup of the SWEAP instrument suit. The large amplitude switchbacks are naturally responsible for large variations of the angular velocity but these measurements shows large positive average values around 40 km/s, as well as negative values that are strongly challenging our understanding of the angular momentum carried by the solar wind. 

In this paper, we attempt to model PSP's observations using a newly developed magneto-hydrodynamic (MHD) model of the solar corona, which take into account the Alfvén waves propagation and dissipation. The model relies on recent modelling strategies, solving, in addition to the classical MHD system, two equations describing the evolution of the Alfvén wave energy densities injected at the lower boundaries \citep[see][for similar approaches]{vanderHolst2010,Sokolov2013}. The structure of the solar magnetic field is then a key input to the model, and we use ADAPT maps \citep{Arge2010}, which combine remote photospheric observations and modelling of the solar magnetic field, to constrain our inner boundary condition. 

As the reader will see, the model is in good agreement with the data. This approach can however only hope to model the large scale averaged quantities measured by PSP. We thus propose further that the main mismatch between the model and the data may be explained by the effects of the dynamics of Alfvénic switchbacks in the solar wind. This interpretation is discussed in the context of the computation of the solar wind open flux and angular momentum.

Section \ref{sec:num} is dedicated to the description of the MHD model. In section \ref{sec:sim}, we show the results of the simulation, trace back the origin of the solar wind plasma measured by PSP, and compare in situ measurements with the plasma parameters interpolated along PSP's trajectory. In section \ref{sec:Alfdyn}, we tackle the main mismatches between the data and the model, and we make the hypothesis that they may be solved by including the Alfvén wave contributions to the large scale solar wind properties. We discuss the limitations of our findings and future prospects in Section \ref{sec:dis}.

\section{Model description}
\label{sec:num}

\subsection{MHD system and source terms}
The MHD model has been developed starting from the PLUTO code \citep{Mignone2007}. The MHD equations are solved in conservative form for the background flow while the contribution of the waves' energy ($\mathcal{E} = \mathcal{E}^+ + \mathcal{E}^-$) and pressure ($p_w = \mathcal{E}/2$) is accounted for \citep{Dewar1970,Jacques1977}. The system can be written:

\begin{equation}
\label{MHD_1}
\frac{\partial}{\partial t} \rho + \nabla \cdot \rho \mathbf{v} = 0,
\end{equation}
\begin{equation}
\label{MHD_2}
\frac{\partial}{\partial t} \rho \mathbf{v} + \nabla \cdot (\rho \mathbf{vv}-\mathbf{BB}+\mathbf{I}p) = - \rho \nabla \Phi,
\end{equation}
\begin{align*}
\label{MHD_3}
\frac{\partial}{\partial t} (E + \mathcal{E} + \rho \Phi)  +& \nabla \cdot [(E+p+\rho \Phi)\mathbf{v}\\-\mathbf{B}(\mathbf{v}\cdot \mathbf{B}) + &\mathbf{v}_g^+ \mathcal{E}^+ + \mathbf{v}_g^- \mathcal{E}^-] = Q, \numberthis
\end{align*}
\begin{equation}
\label{MHD_4}
\frac{\partial}{\partial t} \mathbf{B} + \nabla \cdot (\mathbf{vB}-\mathbf{Bv})=0,
\end{equation}
where $E \equiv  \rho e + \rho v^2/2 + B^2/2$ is the background flow energy, $\mathbf{B}$ is the magnetic field, $\rho$ is the mass density, $\mathbf{v}$ is the velocity field, $p = p_{\mathrm{th}}+ \mathcal{E}/2 + B^2/2$ is the total (thermal, wave and magnetic) pressure, $\mathbf{I}$ is the identity matrix and $\mathbf{v}_g^{\pm} = \mathbf{v} \pm \mathbf{v_A}$ is the group velocity of Alfvén wave packets (see section \ref{subsec:numwaves}).

The system is solved in spherical coordinates $(r,\theta, \varphi)$ and the gravity potential

\begin{equation}
\Phi = - \frac{GM_{\odot}}{r}.
\end{equation}

The source term $Q$ added to the energy equation is made of four components:
\begin{equation}
\label{eq:q}
Q = Q_h +Q_w - Q_c - Q_r.
\end{equation}
The heating $Q_h + Q_w$ is split between two sources, an ad hoc function $Q_h$ and a turbulence term $Q_w$, which will be further described in the next subsection. The ad hoc term

\begin{equation}
Q_h = F_h/H \left(\frac{R_{\odot}}{r} \right)^2 \exp{ \left(-\frac{r-R_{\odot}}{H}\right)},
\label{eq:adhoc}
\end{equation}
with $H \sim 1 R_{\odot}$, the heating scale-height, and $F_h$ the energy flux from the photosphere \citep[in erg.cm$^{-2}$s$^{-1}$, see e.g.,][]{Grappin2010}. 

We then use an optically thin radiation cooling prescription,

\begin{equation}
Q_r = n^2 \Lambda (T),
\end{equation}
with $n$ the electron density and $T$ the electron temperature. $\Lambda(T)$ follows the prescription of \citet{Athay1986}. The thermal conduction is written 

\begin{equation}
Q_c = \nabla \cdot (\alpha \mathbf{q}_s + (1-\alpha) \mathbf{q}_p),
\end{equation}
where $\mathbf{q}_s = -\kappa_0 T^{5/2} \nabla T$ is the usual Spitzer-Härm collisional thermal conduction with $\kappa_0=9 \times 10^{-7}$ cgs, and $\mathbf{q}_p = 3/2 p_{\mathrm{th}} \mathbf{v}_e$ is the electron collisionless heat flux described in \citet{Hollweg1986}. The coefficient $\alpha = 1/(1+(r-R_{\odot})^4/(r_{\mathrm{coll}}-R_{\odot})^4)$ creates a smooth transition between the two regimes at a characteristic height of $r_{\mathrm{coll}} = 5 R_{\odot}$. The system is closed by an ideal equation of state relating the internal energy and the thermal pressure, 

\begin{equation}
\rho e  = \frac{p_{\mathrm{th}}}{\gamma-1},
\end{equation}
with $\gamma = 5/3$, the ratio of specific heat for a fully ionized hydrogen gas. The equations are solved using an improved Harten, Lax, van Leer Riemann solver \citep[HLL, see][]{Einfeldt1988} and a parabolic reconstruction method with minmod slope limiter. $\nabla \cdot \mathbf{B} = 0$ is ensured through the hyperbolic divergence cleaning method \citep{Dedner2002}. To this system we add two equations of wave energy propagation and dissipation which give the terms $\mathcal{E} = \mathcal{E}^+ + \mathcal{E}^-$and $Q_w^\pm$ and which are described in the following subsection.

\subsection{Wave propagation, dissipation and heating}
\label{subsec:numwaves}

We propagate two populations of parallel and anti-parallel Alfvén waves from the boundary conditions. The Els\"asser variables are defined as follows:

\begin{equation}
\mathbf{z}^\pm = \delta \mathbf{v} \mp \mathrm{sign}(B_r) \frac{\delta \mathbf{b}}{\sqrt{\mu_0 \rho}},
\end{equation}
so that the sign + (-), corresponds to the forward wave in a + (-) field polarity. The wave energy propagation follows the WKB theory \citep[see][]{AlazrakiCouturier1971,Belcher1971,Hollweg1974,TuMarsch1993,TuMarsch1995}. These equations read:

\begin{equation}
\frac{\partial \mathcal{E}^\pm}{\partial t} + \nabla \cdot \left( [\mathbf{v} \pm \mathbf{v_A}] \mathcal{E}^\pm \right) = -\frac{\mathcal{E}^\pm}{2} \nabla \cdot \mathbf{v}  - Q_w^\pm,
\end{equation}
where

\begin{equation}
\mathcal{E}^\pm =  \rho \frac{|z^\pm|^2}{4}
\end{equation}
is the wave energy density for each wave population and

\begin{equation}
Q_w = Q_w^+ + Q_w^-,
\end{equation}
where each term 

\begin{equation}
Q_w^\pm = \mathcal{E^\pm} \frac{|z^{\mp}|}{2 \lambda}=\rho  |z^\pm|^2 \frac{|z^\mp|}{8 \lambda}.
\label{eq:kol_ph}
\end{equation}

This term follows the Kolmogorov phenomenology assuming a scale-invariant cascade of the Alfvén wave energy and a complete separation of the injection scale and the dissipation scale. The dissipation length scale $\lambda$ is thus set according to the large scale correlation length of the Alfvén waves, which is usually close to the size of super granules in the low corona and increases with the square root of the magnetic field, i.e. the width of the flux tube, $\lambda = \lambda_{\odot} /\sqrt{B}$. This approach, while not describing in details the cascading process and the dissipation, is a good approximation for such a large scale study, as we shall see later in this work. 

Closed loops, where the magnetic field confines the coronal plasma, and open regions are created self-consistently while the code relaxes to a steady state. In closed loops, the dissipation is mostly obtained by the interaction of the two counter-propagating waves population. In order to have turbulent dissipation in open regions as well, we set a small constant reflection coefficient to create an inward wave population which is instantly dissipated. The dissipation terms hence read:

\begin{equation}
Q_w^\pm = \frac{\rho}{8}  \frac{|z^\pm|^2}{\lambda} (\mathcal{R} |z^\pm| + |z^\mp|),
\label{eq:kol_ph2}
\end{equation}
where $\mathcal{R}=0.1$, which yields a heating rate consistent with incompressible turbulence studies \citep[see for instance][]{VerdiniVelli2007,ChandranHollweg2009,vanderHolst2010}. The parameters of the simulations are hence essentially $\delta v_{\odot} = z^{\pm}_{\odot}/2$, $\lambda_{\odot}$ and the photospheric magnetic field, which is set as a boundary condition using observations.

\section{Simulation of PSP encounter 1}
\label{sec:sim}

\subsection{Simulation parameters}

In order to compare the  numerical MHD simulations with the measurements, we chose to use Air Force Data Assimilative Photospheric Flux Transport (ADAPT) map \citep{Arge2010} of the solar photospheric magnetic field on November 6th, 2018 at 12:00 UTC\footnote{https://www.nso.edu/data/nisp-data/adapt-maps/}. The map is first projected on a spherical harmonics decomposition up to an order $\ell=15$. This procedures reduces the amplitude of the radial field from photospheric levels ($> 100$ G) to coronal levels (a few G). The simulation is performed on a grid of $224 \times 96 \times 192$ points in $r$, $\theta$ and $\phi$ respectively. The grids in the angular directions are uniform, while the grid in the radial direction is stretched from the surface (where the highest resolution is $dr = 0.01 R_{\odot}$ at the bottom boundary) to $20 R_{\odot}$ with 128 points and uniform up to $50 R_{\odot}$ with $96$ points. The first radial cell is thus above the transition region, and the domain starts in the low corona, consistently with the input magnetic field. The input transverse velocity is set everywhere to

\begin{equation}
    \delta v_{\odot} = 30 \; \mathrm{km/s},
\end{equation}
so that the total input of Alfvén wave energy is $\langle \rho_{\odot} v_{A,\odot} \delta v_{\odot}^2 \rangle=  \rho_{\odot} \langle v_{A,\odot} \rangle \delta v_{\odot}^2 \approx 0.8 \times 10^{5}$ erg.cm$^{-2}$s$^{-1}$, with $\rho_{\odot} = 5 \times 10^{-16}$ g.cm$^{-3}$ and $\langle B_{\odot} \rangle \approx 1.4$G (the Alfvén wave flux at a given latitude and longitude depends on the precise value of the radial field). An additional small ad hoc flux is used with $F_h = 2 \times 10^{4}$ erg.cm$^{-2}$s$^{-1}$ and a scale height $H=1 R_{\odot}$ (see equation \eqref{eq:adhoc}), to model chromospheric or coronal heating processes that would be different from waves. The total energy input is thus around $1.0 \times 10^{5}$ erg.cm$^{-2}$s$^{-1}$, which is what is required to power the solar wind \citep[see, for example, the appendix of][]{Reville2018}. Finally, the correlation length at the base of the domain is set to $\lambda_{\odot} = 0.022 R_{\odot} \sqrt{\mathrm{G}} \approx 15 000 \; \mathrm{km}\sqrt{\mathrm{G}}$, which is close to the size of supergranules \citep[see][]{VerdiniVelli2007}.

The equations are solved in the rotating frame assuming a period of 25 days, which is the equatorial speed in the solar differential rotation profile, and thus close to what PSP has seen in the vicinity of the ecliptic plane. A steady state is obtained after approximately three Alfvén crossing times. Consequently, we made the choice to run the MHD simulation up to $50 R_{\odot}$, to limit the necessary computing time, already equivalent to 100 thousand CPU hours. This is enough to cover the highest cadence data at perihelion. We then perform an extrapolation to $130 R_{\odot}$, assuming:
\begin{eqnarray}
    n &\propto& r^{-2},\\
    v_r &=& cste, \\
    v_{\theta,\varphi} &\propto & r^{-1},\\
    B_r & \propto & r^{-2}, \\
    B_{\theta,\varphi} &\propto & r^{-1},\\
    T &\propto& r^{-4/3},
\label{eq:extr}
\end{eqnarray}
following a field line along the Parker spiral at the speed given for each latitude and longitude. The wave energy decays accordingly with the WKB theory and we hence assume that the wave heating is negligible beyond $50 R_{\odot}$. The temperature is consequently extrapolated assuming an adiabatic expansion (equation \ref{eq:extr}). The extrapolation extends smoothly the solution, allowing to compute the plasma properties for an extended time interval along PSP's trajectory.

\subsection{Solar wind sources}

\begin{figure*}
    \centering
    \includegraphics[width=6.5in]{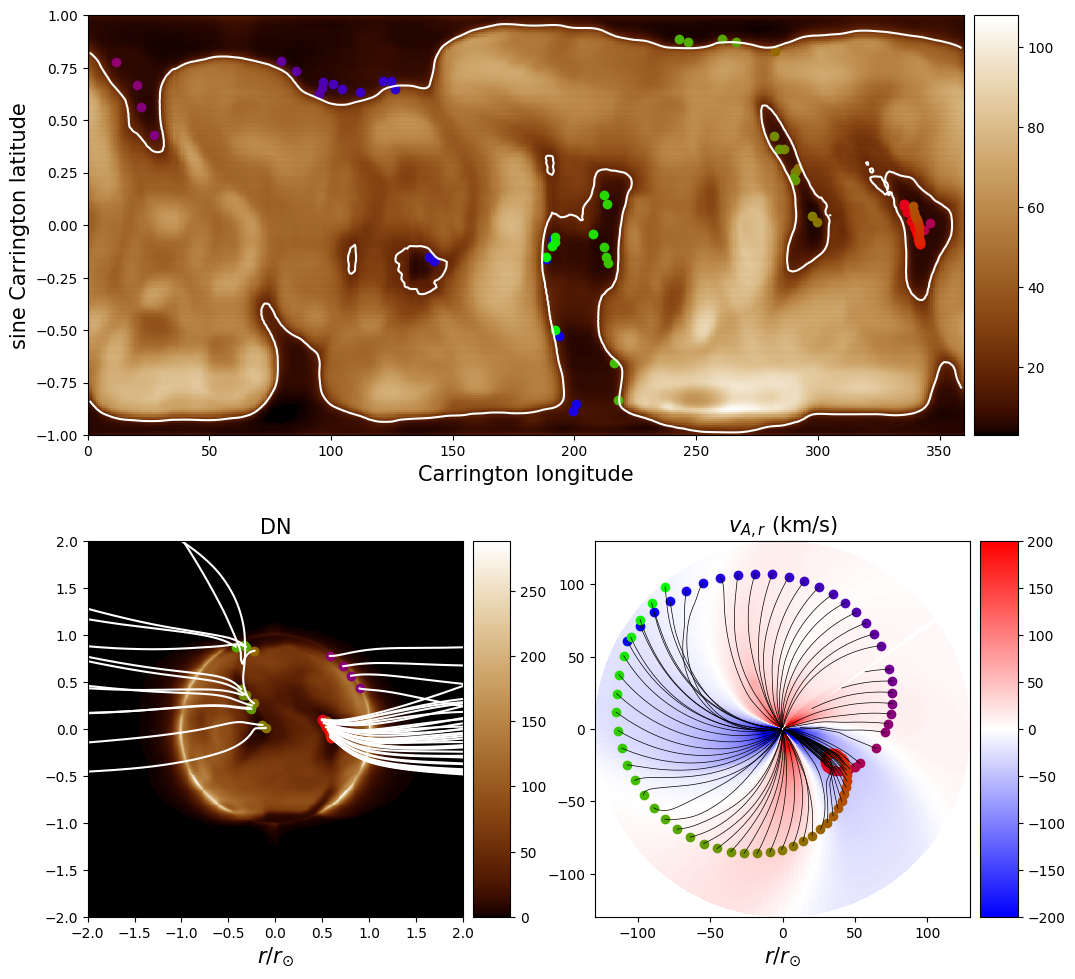}
    \caption{Solar wind footpoints along PSP trajectory from October 15th (blue points) to November 30th (green points). The top panel shows the projection of the source points on a SDO/AIA synthetic synoptic image of the corona at $193$ Angstr\"om. The bottom left panel shows the solar disk with the same technique, viewed from the side, allowing to clearly identify two low-latitude coronal holes connected to PSP at perihelion. In the bottom right panel we see magnetic field lines traced back to the Sun with corresponding colors. The background is the signed radial Alfvén speed in the equatorial plane of the Sun's rotating frame. The color scale is saturated at $200$ km/s to show the polarity changes.}
    \label{fig:sources}
\end{figure*}

The MHD simulation yields the full 3D structure of the corona and thus allows to find the source regions of the plasma measured by PSP. Those sources are identified in Figure \ref{fig:sources}. We selected an interval roughly centered around the perihelion of November 6th: between October 15th and November 30th, and computed the field lines from PSP's position back to the solar surface. Each footpoint has a distinct color and can be identified on all panels in Figure \ref{fig:sources}. In the top and bottom left panels, we synthesize an Extreme UltraViolet (EUV) image of the solar corona from the simulation. We use the response $\mathcal{R}(n,T)$ of the SDO/AIA instrument with the 193 Angstr\"om filter, which yields a photon count, or digital number (DN) produced by each cell of the simulation. The images are then obtained integrating along the line of sight (LOS): 
\begin{equation}
    I = \int_{LOS} n^2 \mathcal{R} (n, T) dl \; \left[\mathrm{DN \; s}^{-1} \; \mathrm{pixel}^{-1}\right].
\end{equation}

The top panel is a synoptic map showing the thermal structure of the corona at all longitudes and thus the coronal holes where the solar wind is thought to come from. Coronal holes are darker regions, here delimited by a simple contour of the DN value (20) on the synoptic map, which provide a good idea of the sources of all open field lines in the simulation. The bottom left panel is the image that would have been obtained by SDO/AIA, provided that the spacecraft could have looked from the side at PSP E1. A similar image could hSzaboave been obtained by STEREO B, if still in operation. Images a few days before perihelion taken by SDO/AIA give similar features. Field lines coming from PSP trajectory are superimposed on the images, showing the 3D structure of the corona. Finally, the bottom right panel shows a cut of the signed, radial, Alfvén speed (hence giving the polarity of the magnetic field) seen from the top, and the field lines traced from PSP's trajectory back to the Sun.

We can thus make the following prediction: during its first encounter, PSP has crossed the heliospheric current sheet four times between October 15th and November 30th. The blue footpoints, before the closest approach, are located in a negative polarity equatorial hole (easily seen from actual images of SDO/AIA on November 6th) and in a positive polarity northern coronal hole. The red/brown points represent the closest approach. We find, accordingly with other studies \citep{Badman2020ApJS,Panasenco2020ApJS}, that the plasma is mostly coming from a region close to the equator of negative polarity. On the way out of perihelion, PSP has measured plasma from an adjacent equatorial coronal hole of positive polarity, identified with the brown/green points. As it will be seen in the next section, our model is actually missing one change of polarity early in the approach phase (blue/purple points), and we think that this mismatch essentially comes from the evolution of the photospheric magnetic field in time, which is not taken into account in our model. However, during the closest approach, where our model is more reliable, PSP has probed plasma coming from successive confined low-latitude regions of the Sun, with well-defined and unified properties. 

\subsection{In situ comparison}

\begin{figure*}
    \centering
    \includegraphics[width=7in]{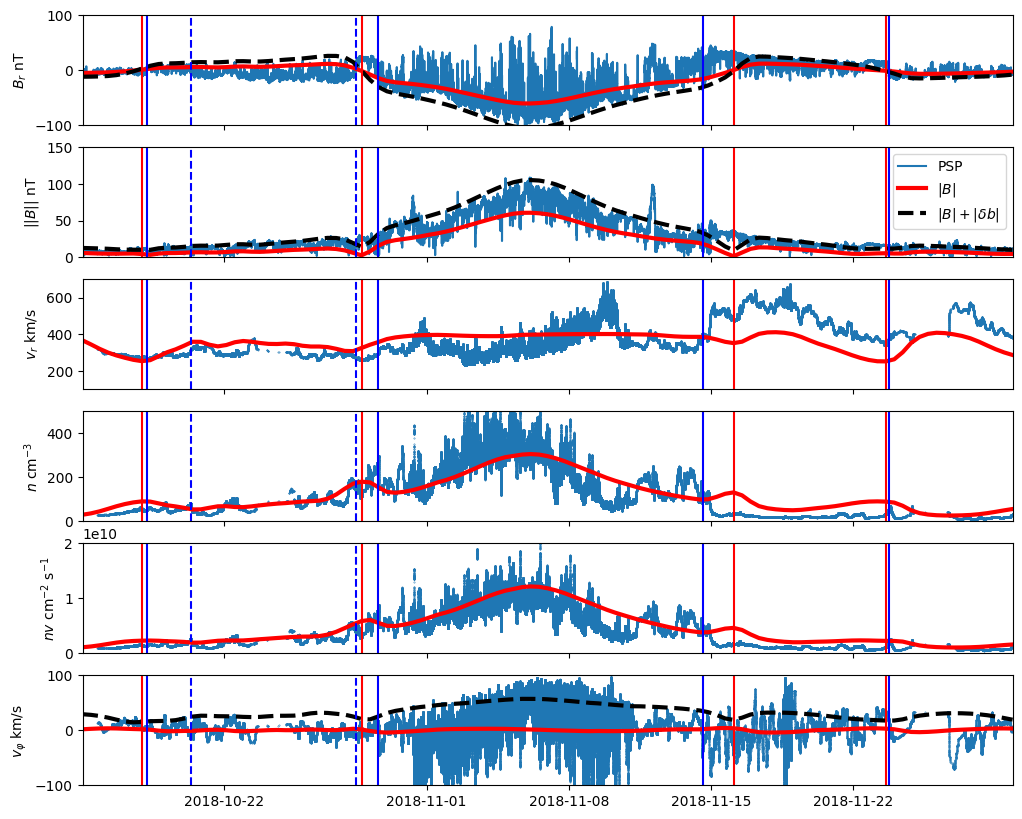}
    \caption{Comparison between in situ magnetic fields and plasma measurements made by Parker Solar Probe and the 3D MHD simulation. The blue lines (or dots) represents measurements, while thick and red curves mark the results from the MHD simulation. Blue and red vertical lines mark HCS crossing in the data and in the simulation respectively.  Dashed black lines are accounting for the Alfvén magnetic and velocity perturbations $\delta b, \delta v$ computed from the wave energy solved by the code (see section \ref{sec:Alfdyn}, and Figure \ref{fig:brav}).}
    \label{fig:dataComp}
\end{figure*}

In Figure \ref{fig:dataComp}, we show the data obtained with the FIELDS and SWEAP instruments on board Parker Solar Probe for the first perihelion, between October 15th and November 30th 2018. The one minute averaged magnetic field data (FIELDS) is shown in the first two panels ($B_r$ and $||\mathbf{B}||$), while the particle moments ($n$ and $\mathbf{v}$, SWEAP/SPC) have been computed with varying cadences depending on PSP's distance to the Sun. The highest cadence is around one second at perihelion, between October 31st and November 10th 2018. During the closest approach, PSP was inside a slow Alfvénic wind region with a globally negative polarity. Switchbacks, i.e. fast reversals of the magnetic field associated with velocity jets, are observed throughout the entire time interval and appear clearly during perihelion in Figure \ref{fig:dataComp}. As shown in \citet{Kasper2019Nature,Bale2019Nature,Tenerani2020ApJS,Horbury2020ApJS}, these jets are Alfvénic, and maintain a high velocity/magnetic field correlation. All vector fields thus show important variations. Finally, the angular velocity field (bottom panel) displays a roughly symmetric profile around perihelion, going on average down to -20 km/s, up to 40 km/s and down to negative values again.

The red profiles in Figure \ref{fig:dataComp} are the results of the interpolation of PSP's trajectory on the simulation (and extrapolation beyond $50 R_{\odot}$). The negative polarity observed during perihelion is recovered in our simulation. The amplitude of the radial field appears about $30$\% lower than the peak of the signal. This trend is also recovered in the total field, for which $B_r$ is the dominant component. The wind speed in the simulation varies between 300 and 400 km/s, and is most of the time in agreement with the data, except for a fast wind event on the way out of perihelion. Density and momentum (forth and fifth panels) agree well with the data. The simulation azimuthal velocity profile is however very flat, with values between $\pm 5$ km/s at the closest approach.

A convenient way to further analyze and compare our simulation results with the data is to compute heliospheric current sheet (HCS) crossings. They are identified with red vertical lines for the simulation and blue lines for the data in each panel of Figure \ref{fig:dataComp}. In the data, many magnetic field reversals are observed, corresponding to switchbacks, and the identification of the HCS crossing can be better asserted with the help of particle measurements (particularly looking at the strahl of the electrons). We use the HCS crossings identified by \citet{Szabo2020ApJS}. The simulation captures most of the HCS crossings except two (which correspond to one switch to a negative polarity region, marked with dashed lines) between October 19th and October 28th.  We observe a one and a half day delay for the HCS crossing on the way out of the perihelion, on November 14th 2pm UTC in the data, and November 16th 2am in the simulation. As shown in the previous section (going from red/brown to green points), this correspond to the switch from a first equatorial coronal hole to a second one. The wind speed coming from the second coronal hole is significantly higher in the data, approaching $600$ km/s, and may be considered as a fast wind component. The mismatch with the simulation could be due to additional wind driving coming from this precise region. Moreover, a stream interaction can be seen from the very sharp wind speed transition observed in the data around November 15th. Hence, this delay may be due to fast/slow wind stream interaction.

\begin{figure}
    \centering
    \includegraphics[width=3.2in]{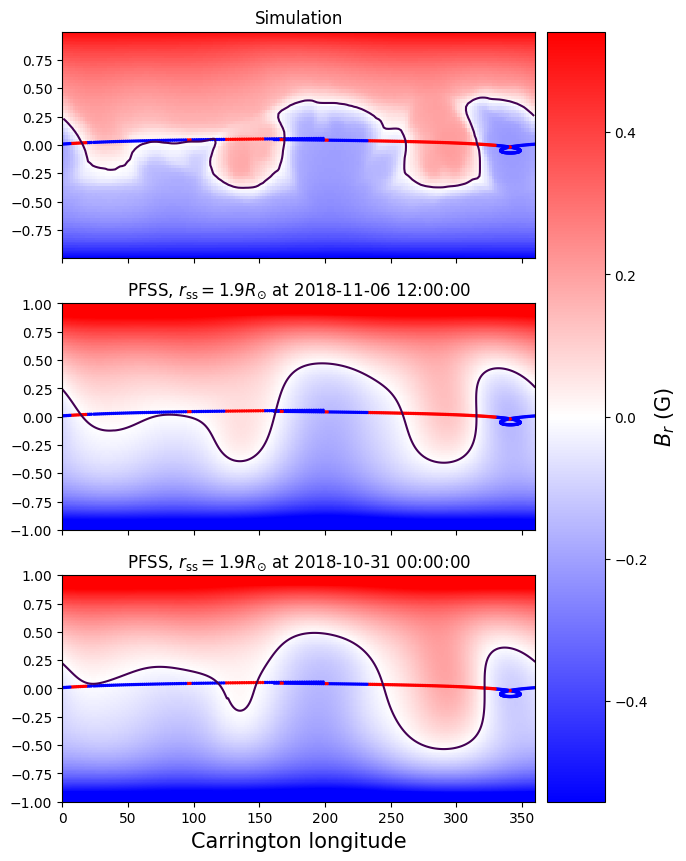}
    \caption{Magnetic field structure at $1.9 R_{\odot}$ in the MHD simulation (upper panel), a PFSS extrapolation using the same magnetic map (middle panel), and another PFSS extrapolation on an October 31st map. The trajectory of the spacecraft between October 15th and November 30th is projected with colors corresponding to the measured polarity. We see that the MHD simulation and the PFSS modeling are close at $1.9 R_{\odot}$ for a given time, and that the earlier map recovers well the polarity change before the perihelion.}
    \label{fig:maptime}
\end{figure}

The dashed lines in Figure \ref{fig:dataComp} represent, as we said, a polarity switch that is not recovered in the simulation. We believe that this can be explained by the evolution of the photospheric magnetic field with time. Our simulation only uses one magnetogram, close to perihelion, in order to save computing time. However, as shown by the study of \citet{Badman2020ApJS}, using time-varying magnetic field maps and potential field source surface models \citep[PFSS,][]{AltschulerNewkirk1969,Schatten1969} can provide a very good match of the HCS crossings. We show in Figure \ref{fig:maptime} the results of the projection in Carrington coordinates of the PSP trajectory on the magnetic field obtained by the simulation, a PFSS model of the map at perihelion (the very same used for the boundary condition of the simulation) and another one taken at a previous time (October 31st, 00:00 UTC). PSP's trajectory projection is accounting for a Parker spiral with a uniform speed of $340$ km/s, which is the average speed observed in the considered time interval. It starts at about 130 degrees of longitude on October 15th and goes to the left. The missing polarity change is obtained with the October 31st map, and it is reasonable to assume that it would have been obtained in the MHD simulation using this map as a boundary condition (with the risk of creating other discrepancies later). It is worth noting here that during perihelion, the solar surface connected to PSP was on the limbs and consequently earlier magnetograms might be more accurate simply because they are directly observed. The source surface radius of the extrapolation in Figure \ref{fig:maptime} is chosen to match the total open flux of the simulation, and we obtain $r_{ss} = 1.9 R_{\odot}$ \citep[see][]{Reville2015b}. Note that the study of \citet{Badman2020ApJS}, indicates that even lower values of the source surface radius provide a better agreement of the polarity changes.  
 
Hence, precise studies of the polarity changes tend to show that over a few weeks, the surface magnetic field of the Sun evolves enough to involve large scale modifications, which can create a mismatch between the data and a single epoch simulation. However, it is probably fair to say that our MHD modelling is able to largely recover the bulk properties of the solar wind observed by PSP during the first perihelion, except for two things: the azimuthal velocity and the amplitude of the radial field. In the following section, we suggest that these observations could be the result of the dynamics of Alfvénic switchbacks.

\section{Alfvén waves dynamics}
\label{sec:Alfdyn}

\subsection{Open flux and parallel wave pressure}

In section \ref{sec:sim}, we stated that the radial magnetic field was lower in the simulation than in the data. The large variability of the observed signal requires to consider things carefully. Switchbacks, or magnetic field reversals, may suggest that the envelope of the signal is the signature of the average or steady coronal magnetic field. If this is true, the measured signal is indeed clearly higher than the radial magnetic field obtained by the simulation (see Figure \ref{fig:dataComp}).

\begin{figure}
    \centering
    \includegraphics[width=3.2in]{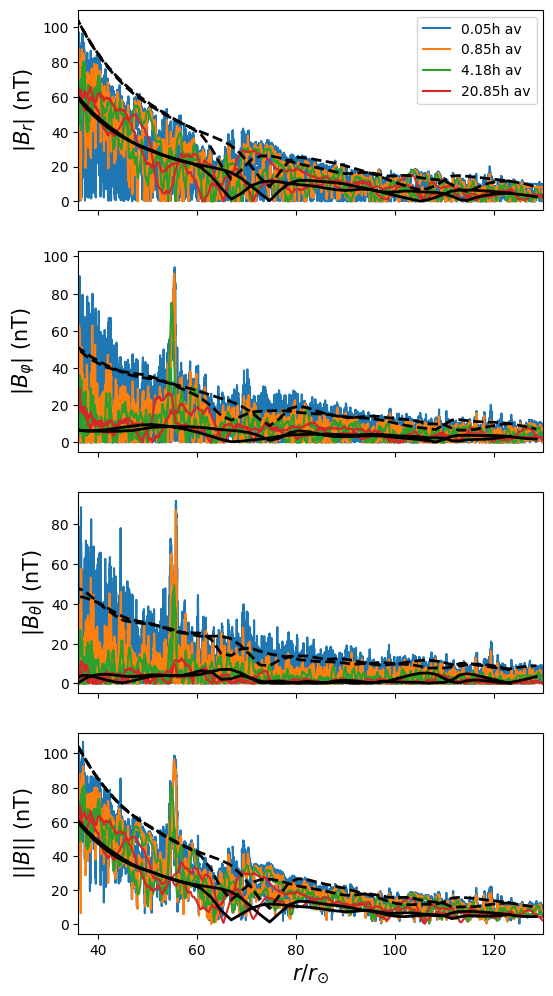}
    \caption{Magnetic field measurements, with various running average timescales, as a function of the radial distance. The largest time average fits fairly well with the radial dependency of the field obtained in the simulation, shown in black. The dashed line illustrates the amplitude of the field when the Alfvén waves are accounted for (see text).}
    \label{fig:brav}
\end{figure}

However, we can obtain a better agreement using usual time averaging approaches. In Figure \ref{fig:brav}, we show the three magnetic field components and the total field, with various running averages and compare this to the simulation results. We find that, at the largest running average presented here ($20.85$ hours, red curves), the agreement for all three components matches the simulation results (in black). This average value is thus significantly lower than the envelope of the perturbations, and it is very clear from the data that these perturbations are non-linear ($\delta B_i \sim \langle B_i \rangle$) and non-transverse ($\delta B_r \sim \delta B_{\theta,\varphi}$). Few studies have addressed the case of large-scale perturbations along the average field direction, but the work of \citep{Hollweg1978a,Hollweg1978b} suggests that the WKB theory could also apply to switchbacks and that the formalism used for our simulation remains valid.

We can thus try to add, in the simulated field measurements, the contribution of the wave population given by the simulation to accelerate the solar wind and heat the corona. The perturbed field can be written 

\begin{equation}
    \delta v^\pm = \frac{\delta b^\pm}{\sqrt{\mu_0 \rho}} = \frac{1}{2}\sqrt{4\mathcal{E}^\pm/\rho},
\end{equation}
assuming an equipartition between the magnetic field and the velocity perturbations. The sum of $B_i + \delta b$ (using the forward wave energy depending on the field polarity) is shown in dashed black in Figure \ref{fig:brav} and Figure \ref{fig:dataComp}. Both curves remarkably match the envelope of the total and radial field signal \citep[note that the peak observed in the vector magnetic field around $50 R_{\odot}$, is a coronal mass ejection that is logically not reproduced by the simulation, see][for more details on this event]{McComas2019Nature}. This essentially means that the Alfvén wave heating scenario used to power the solar wind in the simulation is in agreement with the amplitude of the observed waves and jets along PSP trajectory. Moreover, in the simulation and in the data, the average radial field is around $60$ nT at perihelion, which means that $\langle B_r \rangle$ should be around 1.7 nT at 1 au assuming a $1/r^2$ dependency. This is roughly consistent with observed averages of the open flux during solar minimum, and with OMNI data averaged over October and November 2018: $\langle B_{r,\mathrm{OMNI}} \rangle = 2$ nT. The computed open flux is thus consistent in the simulation and the data, once the perturbations have been removed.

\begin{figure}
    \centering
    \includegraphics[width=3.2in]{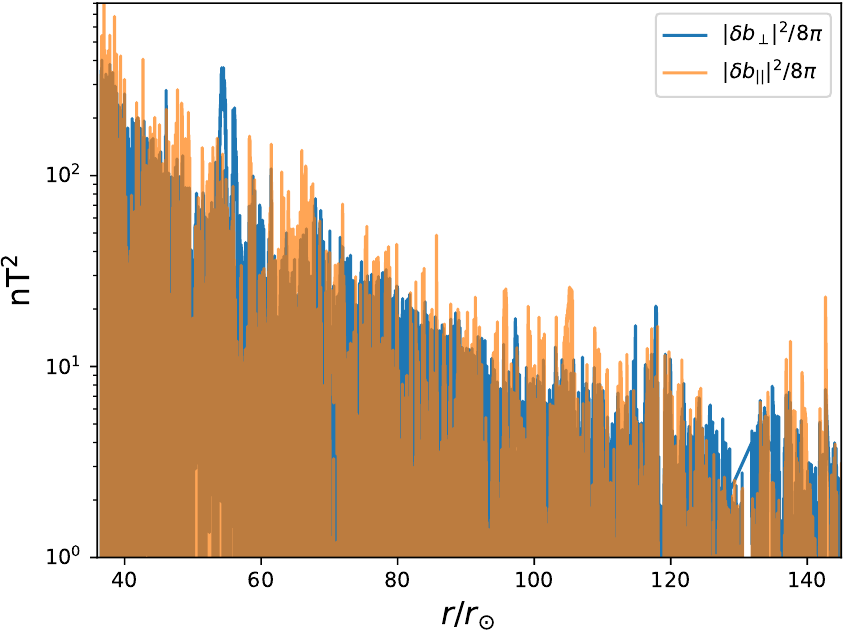}
    \caption{Parallel and perpendicular components of the magnetic perturbations, expressed in terms of magnetic pressure. The average magnetic field direction is obtained with a time average of $20$ hours. The parallel wave pressure component is of the order of the perpendicular component and even higher close to the Sun.}
    \label{fig:parperp}
\end{figure}

Going further with this analysis, we define
\begin{eqnarray}
    \delta \mathbf{b} &=& \mathbf{B}-\langle \mathbf{B} \rangle\\
    \delta b_{||} &=& \delta \mathbf{{b}} \cdot \frac{\langle \mathbf{B} \rangle}{|| \langle \mathbf{B} \rangle ||} ,\\
    \delta \mathbf{b}_{\perp} &=& \delta \mathbf{{b}} - \delta b_{||} \frac{\langle \mathbf{B} \rangle}{|| \langle \mathbf{B} \rangle ||},
    \label{eq:parperp}
\end{eqnarray}
where the averaging operator $\langle \rangle$ is obtained with a running average (convolution) of $20.85$ hours, shown in Figure \ref{fig:brav} to be close to the simulation average fields. In Figure \ref{fig:parperp}, we computed the perturbed parallel and perpendicular components of the magnetic field, and plotted the resulting pressure against the radial distance to the Sun. Note that $\delta b_{||}$ is mostly negative, since it corresponds to magnetic field reversals. We see that close to the Sun, the parallel wave pressure dominates the perpendicular pressure. These measurements have been shown in section \ref{sec:sim} to be associated with one or two source regions at the Sun and contrast strongly with measurements at 1 au. Further away from the Sun, parallel and perpendicular wave pressure remain comparable, with possibly $|\delta b_{||}|^2$ decaying slightly faster than $|\delta b_{\perp}|^2$ (although different distances will correspond to different flux tubes or plasma sources). The study of \citet{Tenerani2020ApJS} shows that switchbacks could survive up to $20-50 R_{\odot}$, in a relatively unperturbed medium. Beyond 1 au or more, they are only observed in very specific conditions, mostly in the fast wind emanating from large coronal holes and a quiet Sun \citep{Neugebauer2013}. This suggests that most switchbacks will unfold during the solar wind expansion, effectively reducing the parallel wave pressure over the perpendicular one.

\subsection{The angular momentum paradox}

\begin{figure*}
    \centering
    \includegraphics[width=6in]{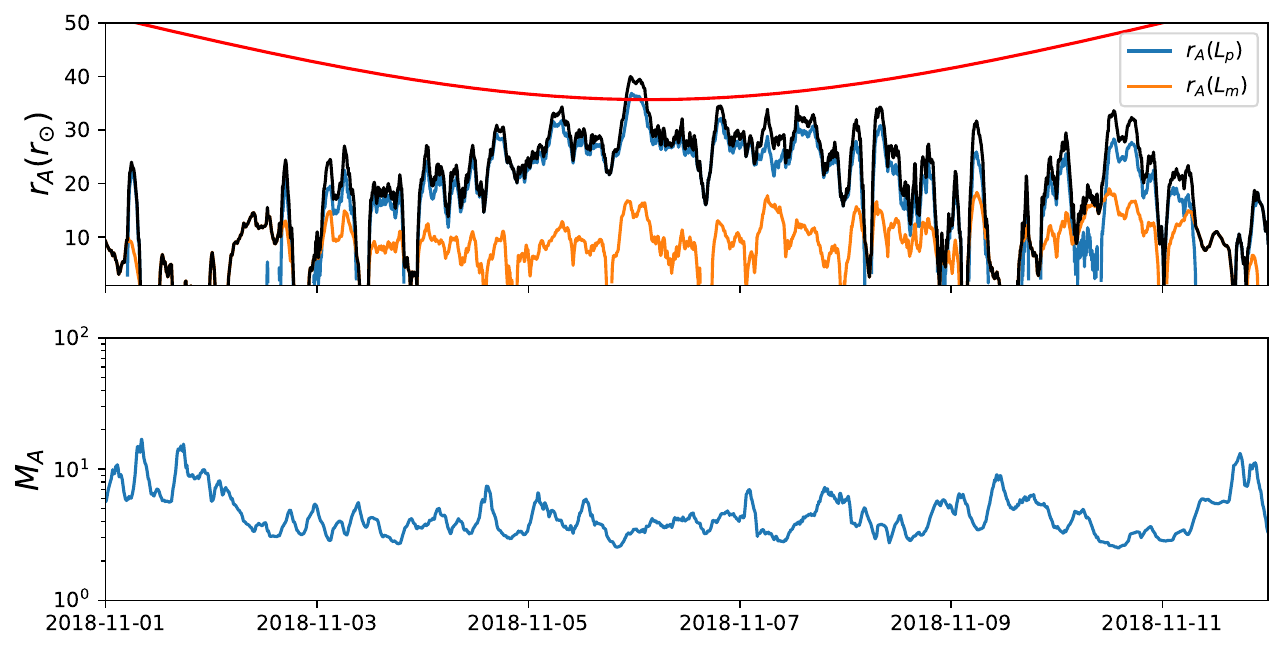}
    \caption{Upper panel : local estimate of the Alfvén radius computed with both terms on equation \eqref{eq:AML}, with the azimuthal velocity in blue, the magnetic stresses in orange and the sum in black. The red line is PSP's distance to the Sun, which crosses the estimated Alfvén point at perihelion. Bottom panel: local Alfvén Mach number, which always stays above one.}
    \label{fig:aml}
\end{figure*}

The Solar Probe Cup (SWEAP/SPC), has revealed an unexpectedly high $v_{\varphi}$ component reaching over 40 km/s on average per second at the perihelion. As shown in Figure \ref{fig:dataComp}, our numerical simulation does not recover these observations. Following the previous section, we added in Figure \ref{fig:dataComp}, the profile of the velocity perturbations to the azimuthal speed obtained in the simulation. The procedure is not as efficient as for the magnetic field data. The peak of the measured tangential velocities is still above the black dashed line, and the average values of observed $v_{\varphi}$ depart clearly from the the average red line of the simulation, which has only a few km/s $\varphi$-velocities at the closest approach. 

Such high angular velocity measurements mean a significant angular momentum of the particles, at least locally, i.e. along the flux tubes crossed by PSP. It has been long known that the angular momentum in the solar wind is strongly related to the Alfvén critical point \citep{WeberDavis1967}. The specific angular momentum along a given field line is a conserved quantity in ideal MHD, and the study of the MHD integral equations along the Parker spiral yields the following result \citep[see, e.g.,][]{Sakurai1985}: 

\begin{equation}
    \Omega_{\odot} r_A^2 = \mathcal{L}_p + \mathcal{L}_m = r v_{\varphi}- r\frac{B_r B_{\varphi}}{4 \pi \rho v_r}.
    \label{eq:AML}
\end{equation}

Hence, the local estimate of the Alfvén radius squared is the sum of two positive terms, one due to the velocity of the particles $\mathcal{L}_p$, and the other associated with the Maxwell stresses $\mathcal{L}_m$ \citep[see][]{Marsch1984}. 

In Figure \ref{fig:aml}, we compute an estimate with each term $r_A (\mathcal{L}_{p,m})$, and the sum $r_A = \sqrt{r_A (\mathcal{L}_p)^2+r_A (\mathcal{L}_m)^2}$ and we compare this to the position of PSP. The estimate computed with the magnetic term, usually thought to be the most reliable, is relatively constant in time and around $10 R_{\odot}$. Using the particles' azimuthal velocity, we reach much higher values, closer to $30 R_{\odot}$. Interestingly, at the very perihelion, the sum of the two terms is higher than the radial distance of PSP to the Sun, during a time interval of about $5$ hours. This measurement occur just after a low frequency magnetic field reversal, and $v_{\varphi}$ measurements go up to $70$ km/s. Around perihelion, PSP's trajectory remains very close to the estimated Alfvén radius while, as shown in the lower panel of Figure \ref{fig:aml}, the Alfvén Mach number $M_A$ has a very flat profile. $M_A$ reaches a minimum of $\sim 4$ at perihelion, far from a subalfvénic regime.

We are hence facing a paradox that calls to revisit equation \eqref{eq:AML}. Strictly speaking, this equation is valid only for an axisymmetric, steady, wind solution. In the simulation, we observe some variation of the angular wind velocity due to azimuthal pressure gradients between slower and faster wind streams. However, as shown in Figure \ref{fig:dataComp}, when compared to the bulk of the observed $v_{\varphi}$ these variations are extremely weak, between $\pm 5$ km/s at most. We thus would like to explore another possibility to solve this paradox. Equation \eqref{eq:AML} is modified when the pressure tensor is no longer scalar:
\begin{equation}
    \Omega_{\odot} r_A^2 = r \left[ v_{\varphi} - \frac{B_r B_{\varphi}}{4 \pi \rho v_r} + \frac{P_{r,\varphi}}{\rho v_r} \right],
    \label{eq:AMLan}
\end{equation}
where $P_{r,\varphi}$ is the $(r,\varphi)$ component of the pressure tensor \citep{Weber1970}. The two first terms on the right hand side are positive (since $B_r B_{\varphi} < 0$), and we thus need $P_{r,\varphi} < 0$, to decrease the whole right hand side term and get an Alfvén critical point compatible with observations. When no waves are present, the pressure tensor is proportional to $p_{\perp} - p_{||}$, and $p_{||}/p_{\perp} > 1$ yields the appropriate behaviour. Accounting for both Alfvén waves and pressure anisotropies we can further write:

\begin{equation}
    P_{r,\varphi} = (p_{_\perp} - p_{||}) f(\mathbf{\langle B \rangle},\delta \mathbf{B}).
\end{equation}
\citet{Hollweg1973} looked at the form of $f$ for purely transverse Alfvén waves and found that the effect of the perturbations is to oppose the effect of larger $p_{||}$. Without going into a full analytical derivation of this term in the case of switchbacks, we can make the hypothesis that perturbations parallel to the magnetic field could inversely strengthen the effect of large parallel over perpendicular pressures. In Figure \ref{fig:parperp}, we see that the parallel magnetic pressure $\delta b_{||}^2 /8\pi$ is higher than the perpendicular pressure close to the Sun, and could consequently help solving this paradox. In the following, we call $\tilde{p}_{||,\perp}$, the pressure tensor components that take into account the contribution of Alfvén waves.

\begin{figure}
    \centering
    \includegraphics[width=3.in]{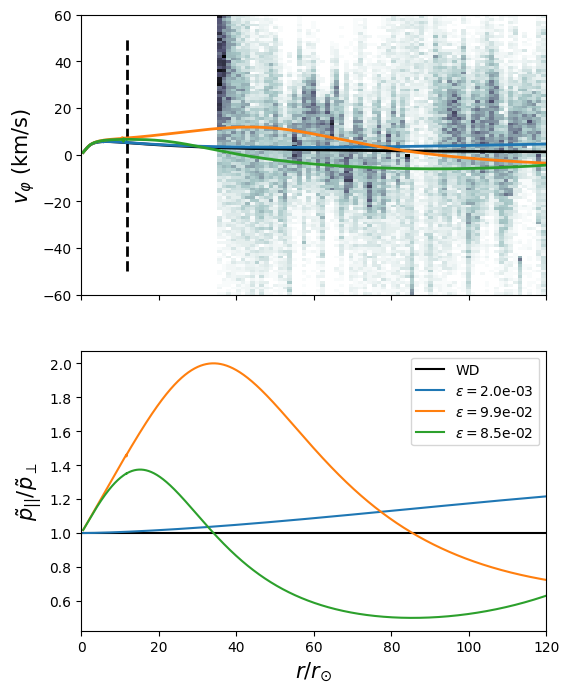}
    \caption{Comparison between the azimuthal velocities measured by SWEAP/SPC (2D histogram in the top panel) with a classical Weber and Davis model (in black) and three models assuming various profiles of pressure anisotropies, which can be seen in the bottom panel. The Alfvén point is almost identical in all models and is shown with the dashed vertical line. The azimuthal velocity increases and decreases (potentially to negative values) with $\tilde{p}_{||}/\tilde{p}_{\perp}$.}
    \label{fig:wd}
\end{figure}

To understand further the effect of anisotropies on the azimuthal velocities, we now look into the simplified analytical model introduced by \citet{WeberDavis1970,Weber1970}. Figure \ref{fig:wd} shows a comparison between the $v_{\varphi}$ measurements and what to expect from several anisotropy profiles. All are based on the \citet{WeberDavis1967} canonical model, assuming a radial field of $2$ nT at $1$ au and a mass flux of $1.4 \times 10^{12}$ g/s, which correspond to the values observed at PSP's perihelion extrapolated to Earth's orbit. The Alfvén radius obtained with this model is located at $11.7 R_{\odot}$, which is close to the average Alfvén radius of the simulation ($\sim 10 R_{\odot}$). Using the classical Weber and Davis model (in black), the angular velocity is around 4 km/s at the closest approach, while observations show a density peak around 40 km/s at perihelion. There is thus a difference of one order of magnitude between the scalar pressure model and the data. The Weber and Davis model also entirely excludes negative angular velocities, that are observed to go down to $-20$ km/s before and after perihelion.

The blue, green and orange curves correspond to the computed $v_{\varphi}$ for various $\tilde{p}_{||}/\tilde{p}_{\perp}$ analytical profiles, which are plotted in the bottom panel of Figure \ref{fig:wd}. In the model of \citet{Weber1970}, the radial components $(v_r, B_r)$ are not affected by the anisotropies and the total angular momentum $\Omega r_A^2$ is only slightly modified by a factor $(1-\epsilon)$. The legend of figure \ref{fig:wd} gives the values of $\epsilon$ (which is an output of the model) for each profile. Increased parallel pressure ($\tilde{p}_{||}/\tilde{p}_{\perp} \sim 2$) can thus decrease the total angular momentum by up to $10$\%. In contrast, the tangential velocities are strongly affected, as the system tries to compensate for an almost constant $r_A$ (see equation \ref{eq:AMLan}). $v_{\varphi}$ thus increase when the parallel to perpendicular pressure ratio is above one, in agreement with the results of \citep{WeberDavis1970,Weber1970}. With a factor two in the pressure anisotropies, we can obtain $v_{\varphi} \sim 13$ km/s at the closest approach, i.e. only a factor $3$ in comparison with the observations. Negative values of $v_{\varphi}$ are obtained when the ratio $\tilde{p}_{||}/\tilde{p}_{\perp}$ goes below one. The orange curve is in good agreement with the data between $50$ and $80 R_{\odot}$. The peak of observed tangential velocity at perihelion remains however difficult to reach. Increasing further the parallel over perpendicular pressure ratio would yield higher $v_{\varphi}$. Strong anistropies could also have a meaningful effect on the poloidal components of the magnetic and velocity fields, violating the assumptions of the simple model used here. The study of \citet{Huang2020} shows that proton anisotropies are compatible with $p_{||}/p_{\perp} \approx 2$ but not much more. However, electron velocity distribution functions are more likely to yield larger parallel than perpendicular pressures \citep[notably through the electron beam, see e.g.][]{Marsch2006} and need to be further studied.

\section{Discussion}
\label{sec:dis}

In this work we have compared the large-scale properties measured by Parker Solar Probe during the first encounter with an MHD numerical simulation of the corona and the solar wind. The agreement is good in general for most of the bulk properties of the solar wind: density, vector magnetic field and radial velocity. 

The code relies on the hypothesis that the hot corona and the solar wind mainly find their origin in Alfvén waves launched from the photosphere. Waves exert a ponderomotive force \citep[or wave pressure, see][]{AlazrakiCouturier1971,Belcher1971} that helps accelerate the solar wind. They also develop a turbulent spectrum and dissipate energy at small scales to heat the corona. The cascading process is not precisely solved in our model, and we use the Kolmogorov phenomenology to compute the heating rate from the injection of energy at the largest scales. This requires the choice of a base dissipation length $\lambda_{\odot}$, which we set close to the scale of super granules, following many previous works on incompressible turbulence \citep[see, e.g.,][]{VerdiniVelli2007, ChandranHollweg2009, PerezChandran2013}. \citet{vanBallegooijenAsgariTarghi2017} have argued that the correlation length should be closer to the size of granules, i.e $\sim 1000$ km, where few km/s transverse motions could be the origin of the Alfvén waves propagating in the corona and the solar wind. They do find, however, that the actual dissipation is about one order of magnitude lower than what is given by phenomenological models. The heating rate between our model and theirs is consequently comparable. Moreover, the funnel expansion in the chromosphere and transition region increases the wave amplitudes and the correlation length by around one order of magnitude, which corresponds to the values used in this work for the low corona.

Naturally, the wave turbulence prescription used here is simplified and could be further improved, for instance by explicitly including the Alfvén wave reflection process \citep[see
][]{Lionello2014,vanDerHolst2014,Usmanov2018} and by trying to account for the compressible nature of the solar wind in the cascading process \citep[see][]{vanBallegooijenAsgariTarghi2016,Shoda2018b,Reville2018,Verdini2019}. The model is nonetheless able to produce an accurate 3D structure of the solar corona. The predictions for the HCS crossings are for the most part very close to the observed data. The agreement is limited by the time evolution of the solar photospheric magnetic field, which we keep fixed to rely on a single simulation. The study of \citet{Badman2020ApJS} shows, using PFSS models, that time-varying magnetograms were an important part of the 3D modelling of the corona. It is also likely that the model could reach a better accuracy for perihelia where the Solar surface connected to PSP is facing Earth and other Doppler instruments able to provide a magnetogram from space (SDO/HMI for example).

At first glance, it might appear that the amplitude of the radial field in the simulation is lower than in the data. However, as shown in section \ref{sec:Alfdyn}, a classical averaging procedure allows matching of the data with simulation results. Moreover, the amplitude of the waves given by the simulation is consistent with the total variation of the radial and transverse magnetic field (see Figure \ref{fig:brav}). The actual open flux is thus lower than what is suggested by the envelope of the radial field data. This argument has been already invoked in \citet{Linker2017}, where the authors were unsuccessfully trying to match coronal models using various photospheric synoptic maps with both in situ measurements of the open flux and EUV coronal observations. This has been since known as the open flux problem. The observed radial magnetic field at 1 au to match was between 1.7 and 2.2 nT (for a different epoch but also around solar minimum), depending on whether folds in the magnetic field were accounted or not. Our study shows that it is the lowest value that steady models should aim for (whether PFSS or MHD) and we are able to fully recover it in our simulation. This good agreement may be due to the magnetic map we used, which involves flux transport modeling and possibly enhances the magnetic field in polar regions \citep{Riley2019b}. A thorough discussion on the open flux problem requires nonetheless a careful comparison of the coronal holes boundaries in the model and in EUV remote sensing measurements. This is left for future works.

As shown further in section \ref{sec:Alfdyn}, switchbacks are fully three dimensional and create a significant perturbation component parallel to the average magnetic field. Interestingly, early works on the solar wind angular momentum have tried to explain the high azimuthal velocities ($\sim 5$ km/s) observed at Earth involving pressure tensor anisotropies \citep{WeberDavis1970,Weber1970}. They showed that larger parallel pressures could raise the azimuthal velocities significantly. These anisotropies were understood as temperature anisotropies only, and transverse waves were actually thought to have an opposite effect on the $v_{\varphi}$ component \citep{Hollweg1973}. However, the large parallel pressure created by the switchbacks could be directly linked to the increase of azimuthal velocities. Using the analytical model of \citet{Weber1970}, we were able to obtain a better agreement with the data, with a significant increase of the azimuthal velocities as well as negative values depending on the anisotropy profile $\tilde{p}_{||}/\tilde{p}_{\perp}$. The model is however not fully consistent and further theoretical studies are necessary to assess whether the solar wind total angular momentum could be affected by strong anisotropies and switchbacks. Anisotropies may also only be a small part of the picture, but such large azimuthal velocity measurements should question the current paradigm for the computation of the solar wind braking, which has been a pending question in solar and stellar physics for more than 50 years. Future data from upcoming encounters of Parker Solar Probe will also provide key information on the properties of switchbacks, pressure anisotropies and particles angular momentum close to the Sun.

\section{Acknowledgements}

This research was supported by the NASA Parker Solar Probe Observatory Scientist grant NNX15AF34G. The FIELDS and SWEAP experiments were developed and are operated under NASA contract NNN06AA01C. This work utilizes data produced collaboratively between AFRL/ADAPT and NSO/NISP. VR wishes to thank A.S. Brun and A. Strugarek for their support and contributions to the code development, notably through PNST, CNES, and IRS SpaceObs fundings. VR also acknowledges funding by the ERC SLOW{\_}\,SOURCE project (SLOW{\_}\,SOURCE - DLV-819189), and thanks A. Rouillard for fruitful discussions. The authors are grateful to A. Mignone and the PLUTO development team. Numerical tasks were performed on the Extreme Science and Engineering Discovery Environment \citep[XSEDE,][]{Xsede2014} SDSC based resource Comet with allocation number AST180027, and GENCI supercomputers (grant 20410133). XSEDE is supported by National Science Foundation grant number ACI-1548562. This study has made use of the NASA Astrophysics Data System. We thank the anonymous referee for providing a detailed report that helped improve the quality of the manuscript. 

\bibliographystyle{yahapj}
\bibliography{./biblio}

\begin{thebibliography}{60}
\providecommand\natexlab[1]{#1}
\providecommand\JournalTitle[1]{#1}

\bibitem[{{Alazraki} \& {Couturier}(1971)}]{AlazrakiCouturier1971}
{Alazraki}, G., \& {Couturier}, P. 1971, \JournalTitle{Astronomy and
  Astrophysics}, 13, 380

\bibitem[{{Altschuler} \& {Newkirk}(1969)}]{AltschulerNewkirk1969}
{Altschuler}, M.~D., \& {Newkirk}, G. 1969,
  \href{http://dx.doi.org/10.1007/BF00145734}{\JournalTitle{\solphys}, 9, 131}

\bibitem[{{Arge} {et~al.}(2010){Arge}, {Henney}, {Koller}, {Compeau}, {Young},
  {MacKenzie}, {Fay}, \& {Harvey}}]{Arge2010}
{Arge}, C.~N., {Henney}, C.~J., {Koller}, J., {et~al.} 2010,
  \href{http://dx.doi.org/10.1063/1.3395870}{in American Institute of Physics
  Conference Series, Vol. 1216, Twelfth International Solar Wind Conference,
  ed. M.~{Maksimovic}, K.~{Issautier}, N.~{Meyer-Vernet}, M.~{Moncuquet}, \&
  F.~{Pantellini}}, 343

\bibitem[{{Athay}(1986)}]{Athay1986}
{Athay}, R.~G. 1986,
  \href{http://dx.doi.org/10.1086/164565}{\JournalTitle{\apj}, 308, 975}

\bibitem[{{Badman} {et~al.}(2020){Badman}, {Bale}, {Mart{\'\i}nez Oliveros},
  {Panasenco}, {Velli}, {Stansby}, {Buitrago-Casas}, {R{\'e}ville}, {Bonnell},
  {Case}, {Dudok de Wit}, {Goetz}, {Harvey}, {Kasper}, {Korreck}, {Larson},
  {Livi}, {MacDowall}, {Malaspina}, {Pulupa}, {Stevens}, \&
  {Whittlesey}}]{Badman2020ApJS}
{Badman}, S.~T., {Bale}, S.~D., {Mart{\'\i}nez Oliveros}, J.~C., {et~al.} 2020,
  \href{http://dx.doi.org/10.3847/1538-4365/ab4da7}{\JournalTitle{\apjs}, 246,
  23}

\bibitem[{Bale {et~al.}(2016)Bale, Goetz, Harvey, Turin, Bonnell,
  Dudok de Wit, Ergun, MacDowall, Pulupa, Andre, Bolton, Bougeret, Bowen,
  Burgess, Cattell, Chandran, Chaston, Chen, Choi, Connerney, Cranmer,
  Diaz-Aguado, Donakowski, Drake, Farrell, Fergeau, Fermin, Fischer, Fox,
  Glaser, Goldstein, Gordon, Hanson, Harris, Hayes, Hinze, Hollweg, Horbury,
  Howard, Hoxie, Jannet, Karlsson, Kasper, Kellogg, Kien, Klimchuk,
  Krasnoselskikh, Krucker, Lynch, Maksimovic, Malaspina, Marker, Martin,
  Martinez-Oliveros, McCauley, McComas, McDonald, Meyer-Vernet, Moncuquet,
  Monson, Mozer, Murphy, Odom, Oliverson, Olson, Parker, Pankow, Phan,
  Quataert, Quinn, Ruplin, Salem, Seitz, Sheppard, Siy, Stevens, Summers,
  Szabo, Timofeeva, Vaivads, Velli, Yehle, Werthimer, \& Wygant}]{Bale2016}
Bale, S.~D., Goetz, K., Harvey, P.~R., {et~al.} 2016,
  \href{http://dx.doi.org/10.1007/s11214-016-0244-5}{\JournalTitle{Space
  Science Reviews}, 204, 49}

\bibitem[{{Bale} {et~al.}(2019){Bale}, {Badman}, {Bonnell}, {Bowen}, {Burgess},
  {Case}, {Cattell}, {Chandran}, {Chaston}, {Chen}, {Drake}, {de Wit},
  {Eastwood}, {Ergun}, {Farrell}, {Fong}, {Goetz}, {Goldstein}, {Goodrich},
  {Harvey}, {Horbury}, {Howes}, {Kasper}, {Kellogg}, {Klimchuk}, {Korreck},
  {Krasnoselskikh}, {Krucker}, {Laker}, {Larson}, {MacDowall}, {Maksimovic},
  {Malaspina}, {Martinez-Oliveros}, {McComas}, {Meyer-Vernet}, {Moncuquet},
  {Mozer}, {Phan}, {Pulupa}, {Raouafi}, {Salem}, {Stansby}, {Stevens}, {Szabo},
  {Velli}, {Woolley}, \& {Wygant}}]{Bale2019Nature}
{Bale}, S.~D., {Badman}, S.~T., {Bonnell}, J.~W., {et~al.} 2019,
  \href{http://dx.doi.org/10.1038/s41586-019-1818-7}{\JournalTitle{\nat}, 576,
  237}

\bibitem[{{Balogh} {et~al.}(1999){Balogh}, {Forsyth}, {Lucek}, {Horbury}, \&
  {Smith}}]{Balogh1999}
{Balogh}, A., {Forsyth}, R.~J., {Lucek}, E.~A., {Horbury}, T.~S., \& {Smith},
  E.~J. 1999,
  \href{http://dx.doi.org/10.1029/1999GL900061}{\JournalTitle{\grl}, 26, 631}

\bibitem[{{Belcher}(1971)}]{Belcher1971}
{Belcher}, J.~W. 1971,
  \href{http://dx.doi.org/10.1086/151105}{\JournalTitle{\apj}, 168, 509}

\bibitem[{{Chandran} \& {Hollweg}(2009)}]{ChandranHollweg2009}
{Chandran}, B. D.~G., \& {Hollweg}, J.~V. 2009,
  \href{http://dx.doi.org/10.1088/0004-637X/707/2/1659}{\JournalTitle{The
  Astrophysical Journal}, 707, 1659}

\bibitem[{{Dedner} {et~al.}(2002){Dedner}, {Kemm}, {Kr{\"o}ner}, {Munz},
  {Schnitzer}, \& {Wesenberg}}]{Dedner2002}
{Dedner}, A., {Kemm}, F., {Kr{\"o}ner}, D., {et~al.} 2002,
  \href{http://dx.doi.org/10.1006/jcph.2001.6961}{\JournalTitle{Journal of
  Computational Physics}, 175, 645}

\bibitem[{Dewar(1970)}]{Dewar1970}
Dewar, R.~L. 1970, \href{http://dx.doi.org/10.1063/1.1692854}{\JournalTitle{The
  Physics of Fluids}, 13, 2710}

\bibitem[{{Einfeldt}(1988)}]{Einfeldt1988}
{Einfeldt}, B. 1988,
  \href{http://dx.doi.org/10.1137/0725021}{\JournalTitle{SIAM Journal on
  Numerical Analysis}, 25, 294}

\bibitem[{{Fox} {et~al.}(2016){Fox}, {Velli}, {Bale}, {Decker}, {Driesman},
  {Howard}, {Kasper}, {Kinnison}, {Kusterer}, {Lario}, {Lockwood}, {McComas},
  {Raouafi}, \& {Szabo}}]{FoxPSP2016}
{Fox}, N.~J., {Velli}, M.~C., {Bale}, S.~D., {et~al.} 2016,
  \href{http://dx.doi.org/10.1007/s11214-015-0211-6}{\JournalTitle{\ssr}, 204,
  7}

\bibitem[{{Grappin} {et~al.}(2010){Grappin}, {L{\'e}orat}, {Leygnac}, \&
  {Pinto}}]{Grappin2010}
{Grappin}, R., {L{\'e}orat}, J., {Leygnac}, S., \& {Pinto}, R. 2010,
  \href{http://dx.doi.org/10.1063/1.3395848}{in American Institute of Physics
  Conference Series, Vol. 1216, Twelfth International Solar Wind Conference,
  ed. M.~{Maksimovic}, K.~{Issautier}, N.~{Meyer-Vernet}, M.~{Moncuquet}, \&
  F.~{Pantellini}}, 24

\bibitem[{{Hazra} {et~al.}(2021){Hazra}, {R{\'e}ville}, {Perri}, {Strugarek},
  {Brun}, \& {Buchlin}}]{Hazra2021}
{Hazra}, S., {R{\'e}ville}, V., {Perri}, B., {et~al.} 2021,
  \href{http://dx.doi.org/10.3847/1538-4357/abe12e}{\JournalTitle{\apj}, 910,
  90}

\bibitem[{{Hollweg}(1973)}]{Hollweg1973}
{Hollweg}, J.~V. 1973,
  \href{http://dx.doi.org/10.1029/JA078i019p03643}{\JournalTitle{\jgr}, 78,
  3643}

\bibitem[{{Hollweg}(1974)}]{Hollweg1974}
---. 1974,
  \href{http://dx.doi.org/10.1029/JA079i010p01539}{\JournalTitle{\jgr}, 79,
  1539}

\bibitem[{{Hollweg}(1978{\natexlab{a}})}]{Hollweg1978b}
---. 1978{\natexlab{a}},
  \href{http://dx.doi.org/10.1029/JA083iA02p00563}{\JournalTitle{\jgr}, 83,
  563}

\bibitem[{{Hollweg}(1978{\natexlab{b}})}]{Hollweg1978a}
---. 1978{\natexlab{b}},
  \href{http://dx.doi.org/10.1007/BF00152474}{\JournalTitle{\solphys}, 56, 305}

\bibitem[{{Hollweg}(1986)}]{Hollweg1986}
---. 1986,
  \href{http://dx.doi.org/10.1029/JA091iA04p04111}{\JournalTitle{\jgr}, 91,
  4111}

\bibitem[{{Horbury} {et~al.}(2020){Horbury}, {Woolley}, {Laker}, {Matteini},
  {Eastwood}, {Bale}, {Velli}, {Chandran}, {Phan}, {Raouafi}, {Goetz},
  {Harvey}, {Pulupa}, {Klein}, {Dudok de Wit}, {Kasper}, {Korreck}, {Case},
  {Stevens}, {Whittlesey}, {Larson}, {MacDowall}, {Malaspina}, \&
  {Livi}}]{Horbury2020ApJS}
{Horbury}, T.~S., {Woolley}, T., {Laker}, R., {et~al.} 2020,
  \href{http://dx.doi.org/10.3847/1538-4365/ab5b15}{\JournalTitle{\apjs}, 246,
  45}

\bibitem[{{Huang} {et~al.}(2020){Huang}, {Kasper}, {Vech}, {Klein}, {Stevens},
  {Martinovi{\'c}}, {Alterman}, {{\v{D}}urovcov{\'a}}, {Paulson}, {Maruca},
  {Qudsi}, {Case}, {Korreck}, {Jian}, {Velli}, {Lavraud}, {Hegedus}, {Bert},
  {Holmes}, {Bale}, {Larson}, {Livi}, {Whittlesey}, {Pulupa}, {MacDowall},
  {Malaspina}, {Bonnell}, {Harvey}, {Goetz}, \& {Dudok de Wit}}]{Huang2020}
{Huang}, J., {Kasper}, J.~C., {Vech}, D., {et~al.} 2020,
  \href{http://dx.doi.org/10.3847/1538-4365/ab74e0}{\JournalTitle{\apjs}, 246,
  70}

\bibitem[{{Jacques}(1977)}]{Jacques1977}
{Jacques}, S.~A. 1977,
  \href{http://dx.doi.org/10.1086/155430}{\JournalTitle{\apj}, 215, 942}

\bibitem[{Kasper {et~al.}(2016)Kasper, Abiad, Austin, Balat-Pichelin, Bale,
  Belcher, Berg, Bergner, Berthomier, Bookbinder, Brodu, Caldwell, Case,
  Chandran, Cheimets, Cirtain, Cranmer, Curtis, Daigneau, Dalton, Dasgupta,
  DeTomaso, Diaz-Aguado, Djordjevic, Donaskowski, Effinger, Florinski, Fox,
  Freeman, Gallagher, Gary, Gauron, Gates, Goldstein, Golub, Gordon, Gurnee,
  Guth, Halekas, Hatch, Heerikuisen, Ho, Hu, Johnson, Jordan, Korreck, Larson,
  Lazarus, Li, Livi, Ludlam, Maksimovic, McFadden, Marchant, Maruca, McComas,
  Messina, Mercer, Park, Peddie, Pogorelov, Reinhart, Richardson, Robinson,
  Rosen, Skoug, Slagle, Steinberg, Stevens, Szabo, Taylor, Tiu, Turin, Velli,
  Webb, Whittlesey, Wright, Wu, \& Zank}]{Kasper2016}
Kasper, J.~C., Abiad, R., Austin, G., {et~al.} 2016,
  \href{http://dx.doi.org/10.1007/s11214-015-0206-3}{\JournalTitle{Space
  Science Reviews}, 204, 131}

\bibitem[{{Kasper} {et~al.}(2019){Kasper}, {Bale}, {Belcher}, {Berthomier},
  {Case}, {Chandran}, {Curtis}, {Gallagher}, {Gary}, {Golub}, {Halekas}, {Ho},
  {Horbury}, {Hu}, {Huang}, {Klein}, {Korreck}, {Larson}, {Livi}, {Maruca},
  {Lavraud}, {Louarn}, {Maksimovic}, {Martinovic}, {McGinnis}, {Pogorelov},
  {Richardson}, {Skoug}, {Steinberg}, {Stevens}, {Szabo}, {Velli},
  {Whittlesey}, {Wright}, {Zank}, {MacDowall}, {McComas}, {McNutt}, {Pulupa},
  {Raouafi}, \& {Schwadron}}]{Kasper2019Nature}
{Kasper}, J.~C., {Bale}, S.~D., {Belcher}, J.~W., {et~al.} 2019,
  \href{http://dx.doi.org/10.1038/s41586-019-1813-z}{\JournalTitle{\nat}, 576,
  228}

\bibitem[{{Linker} {et~al.}(2017){Linker}, {Caplan}, {Downs}, {Riley}, {Mikic},
  {Lionello}, {Henney}, {Arge}, {Liu}, {Derosa}, {Yeates}, \&
  {Owens}}]{Linker2017}
{Linker}, J.~A., {Caplan}, R.~M., {Downs}, C., {et~al.} 2017,
  \href{http://dx.doi.org/10.3847/1538-4357/aa8a70}{\JournalTitle{\apj}, 848,
  70}

\bibitem[{{Lionello} {et~al.}(2014){Lionello}, {Velli}, {Downs}, {Linker}, \&
  {Miki{\'c}}}]{Lionello2014}
{Lionello}, R., {Velli}, M., {Downs}, C., {Linker}, J.~A., \& {Miki{\'c}}, Z.
  2014,
  \href{http://dx.doi.org/10.1088/0004-637X/796/2/111}{\JournalTitle{\apj},
  796, 111}

\bibitem[{Marsch(2006)}]{Marsch2006}
Marsch, E. 2006,
  \href{http://dx.doi.org/10.12942/lrsp-2006-1}{\JournalTitle{Living Reviews in
  Solar Physics}, 3, 1}

\bibitem[{{Marsch} \& {Richter}(1984)}]{Marsch1984}
{Marsch}, E., \& {Richter}, A.~K. 1984,
  \href{http://dx.doi.org/10.1029/JA089iA07p05386}{\JournalTitle{\jgr}, 89,
  5386}

\bibitem[{McComas {et~al.}(2016)McComas, Alexander, Angold, Bale, Beebe,
  Birdwell, Boyle, Burgum, Burnham, Christian, Cook, Cooper, Cummings, Davis,
  Desai, Dickinson, Dirks, Do, Fox, Giacalone, Gold, Gurnee, Hayes, Hill,
  Kasper, Kecman, Klemic, Krimigis, Labrador, Layman, Leske, Livi, Matthaeus,
  McNutt, Mewaldt, Mitchell, Nelson, Parker, Rankin, Roelof, Schwadron,
  Seifert, Shuman, Stokes, Stone, Vandegriff, Velli, von Rosenvinge, Weidner,
  Wiedenbeck, \& Wilson}]{McComas2016}
McComas, D.~J., Alexander, N., Angold, N., {et~al.} 2016,
  \href{http://dx.doi.org/10.1007/s11214-014-0059-1}{\JournalTitle{Space
  Science Reviews}, 204, 187}

\bibitem[{{McComas} {et~al.}(2019){McComas}, {Christian}, {Cohen}, {Cummings},
  {Davis}, {Desai}, {Giacalone}, {Hill}, {Joyce}, {Krimigis}, {Labrador},
  {Leske}, {Malandraki}, {Matthaeus}, {McNutt}, {Mewaldt}, {Mitchell},
  {Posner}, {Rankin}, {Roelof}, {Schwadron}, {Stone}, {Szalay}, {Wiedenbeck},
  {Bale}, {Kasper}, {Case}, {Korreck}, {MacDowall}, {Pulupa}, {Stevens}, \&
  {Rouillard}}]{McComas2019Nature}
{McComas}, D.~J., {Christian}, E.~R., {Cohen}, C.~M.~S., {et~al.} 2019,
  \href{http://dx.doi.org/10.1038/s41586-019-1811-1}{\JournalTitle{\nat}, 576,
  223}

\bibitem[{{Mignone} {et~al.}(2007){Mignone}, {Bodo}, {Massaglia}, {Matsakos},
  {Tesileanu}, {Zanni}, \& {Ferrari}}]{Mignone2007}
{Mignone}, A., {Bodo}, G., {Massaglia}, S., {et~al.} 2007,
  \href{http://dx.doi.org/10.1086/513316}{\JournalTitle{\apjs}, 170, 228}

\bibitem[{{Neugebauer} \& {Goldstein}(2013)}]{Neugebauer2013}
{Neugebauer}, M., \& {Goldstein}, B.~E. 2013,
  \href{http://dx.doi.org/10.1063/1.4810986}{in American Institute of Physics
  Conference Series, Vol. 1539, Solar Wind 13, ed. G.~P. {Zank}, J.~{Borovsky},
  R.~{Bruno}, J.~{Cirtain}, S.~{Cranmer}, H.~{Elliott}, J.~{Giacalone},
  W.~{Gonzalez}, G.~{Li}, E.~{Marsch}, E.~{Moebius}, N.~{Pogorelov},
  J.~{Spann}, \& O.~{Verkhoglyadova}}, 46

\bibitem[{{Panasenco} {et~al.}(2020){Panasenco}, {Velli}, {D'Amicis}, {Shi},
  {R{\'e}ville}, {Bale}, {Badman}, {Kasper}, {Korreck}, {Bonnell}, {Wit},
  {Goetz}, {Harvey}, {MacDowall}, {Malaspina}, {Pulupa}, {Case}, {Larson},
  {Livi}, {Stevens}, \& {Whittlesey}}]{Panasenco2020ApJS}
{Panasenco}, O., {Velli}, M., {D'Amicis}, R., {et~al.} 2020,
  \href{http://dx.doi.org/10.3847/1538-4365/ab61f4}{\JournalTitle{\apjs}, 246,
  54}

\bibitem[{{Perez} \& {Chandran}(2013)}]{PerezChandran2013}
{Perez}, J.~C., \& {Chandran}, B.~D.~G. 2013,
  \href{http://dx.doi.org/10.1088/0004-637X/776/2/124}{\JournalTitle{\apj},
  776, 124}

\bibitem[{{R{\'e}ville} {et~al.}(2015){R{\'e}ville}, {Brun}, {Strugarek},
  {Matt}, {Bouvier}, {Folsom}, \& {Petit}}]{Reville2015b}
{R{\'e}ville}, V., {Brun}, A.~S., {Strugarek}, A., {et~al.} 2015,
  \href{http://dx.doi.org/10.1088/0004-637X/814/2/99}{\JournalTitle{\apj}, 814,
  99}

\bibitem[{{R{\'e}ville} {et~al.}(2018){R{\'e}ville}, {Tenerani}, \&
  {Velli}}]{Reville2018}
{R{\'e}ville}, V., {Tenerani}, A., \& {Velli}, M. 2018,
  \href{http://dx.doi.org/10.3847/1538-4357/aadb8f}{\JournalTitle{\apj}, 866,
  38}

\bibitem[{{R{\'e}ville} {et~al.}(2020){R{\'e}ville}, {Velli}, {Rouillard},
  {Lavraud}, {Tenerani}, {Shi}, \& {Strugarek}}]{Reville2020ApJL}
{R{\'e}ville}, V., {Velli}, M., {Rouillard}, A.~P., {et~al.} 2020,
  \href{http://dx.doi.org/10.3847/2041-8213/ab911d}{\JournalTitle{\apjl}, 895,
  L20}

\bibitem[{Riley {et~al.}(2019)Riley, Linker, Mikic, Caplan, Downs, \&
  Thumm}]{Riley2019b}
Riley, P., Linker, J.~A., Mikic, Z., {et~al.} 2019,
  \href{http://dx.doi.org/10.3847/1538-4357/ab3a98}{\JournalTitle{The
  Astrophysical Journal}, 884, 18}

\bibitem[{{Sakurai}(1985)}]{Sakurai1985}
{Sakurai}, T. 1985, \JournalTitle{\aap}, 152, 121

\bibitem[{{Schatten} {et~al.}(1969){Schatten}, {Wilcox}, \&
  {Ness}}]{Schatten1969}
{Schatten}, K.~H., {Wilcox}, J.~M., \& {Ness}, N.~F. 1969,
  \href{http://dx.doi.org/10.1007/BF00146478}{\JournalTitle{\solphys}, 6, 442}

\bibitem[{{Shoda} {et~al.}(2018){Shoda}, {Yokoyama}, \& {Suzuki}}]{Shoda2018b}
{Shoda}, M., {Yokoyama}, T., \& {Suzuki}, T.~K. 2018,
  \href{http://dx.doi.org/10.3847/1538-4357/aac218}{\JournalTitle{\apj}, 860,
  17}

\bibitem[{{Sokolov} {et~al.}(2013){Sokolov}, {van der Holst}, {Oran}, {Downs},
  {Roussev}, {Jin}, {Manchester}, {Evans}, \& {Gombosi}}]{Sokolov2013}
{Sokolov}, I.~V., {van der Holst}, B., {Oran}, R., {et~al.} 2013,
  \href{http://dx.doi.org/10.1088/0004-637X/764/1/23}{\JournalTitle{\apj}, 764,
  23}

\bibitem[{{Szabo} {et~al.}(2020){Szabo}, {Larson}, {Whittlesey}, {Stevens},
  {Lavraud}, {Phan}, {Wallace}, {Jones-Mecholsky}, {Arge}, {Badman},
  {Odstrcil}, {Pogorelov}, {Kim}, {Riley}, {Henney}, {Bale}, {Bonnell}, {Case},
  {Dudok de Wit}, {Goetz}, {Harvey}, {Kasper}, {Korreck}, {Koval}, {Livi},
  {MacDowall}, {Malaspina}, \& {Pulupa}}]{Szabo2020ApJS}
{Szabo}, A., {Larson}, D., {Whittlesey}, P., {et~al.} 2020,
  \href{http://dx.doi.org/10.3847/1538-4365/ab5dac}{\JournalTitle{\apjs}, 246,
  47}

\bibitem[{{Tenerani} {et~al.}(2020){Tenerani}, {Velli}, {Matteini},
  {R{\'e}ville}, {Shi}, {Bale}, {Kasper}, {Bonnell}, {Case}, {de Wit}, {Goetz},
  {Harvey}, {Klein}, {Korreck}, {Larson}, {Livi}, {MacDowall}, {Malaspina},
  {Pulupa}, {Stevens}, \& {Whittlesey}}]{Tenerani2020ApJS}
{Tenerani}, A., {Velli}, M., {Matteini}, L., {et~al.} 2020,
  \href{http://dx.doi.org/10.3847/1538-4365/ab53e1}{\JournalTitle{\apjs}, 246,
  32}

\bibitem[{Towns {et~al.}(2014)Towns, Cockerill, Dahan, Foster, Gaither,
  Grimshaw, Hazlewood, Lathrop, Lifka, Peterson, Roskies, Scott, \&
  Wilkins-Diehr}]{Xsede2014}
Towns, J., Cockerill, T., Dahan, M., {et~al.} 2014,
  \href{http://dx.doi.org/10.1109/MCSE.2014.80}{\JournalTitle{Computing in
  Science \& Engineering}, 16, 62}

\bibitem[{{Tu} \& {Marsch}(1993)}]{TuMarsch1993}
{Tu}, C.-Y., \& {Marsch}, E. 1993,
  \href{http://dx.doi.org/10.1029/92JA01947}{\JournalTitle{\jgr}, 98, 1257}

\bibitem[{{Tu} \& {Marsch}(1995)}]{TuMarsch1995}
---. 1995, \href{http://dx.doi.org/10.1007/BF00748891}{\JournalTitle{\ssr}, 73,
  1}

\bibitem[{{Usmanov} {et~al.}(2018){Usmanov}, {Matthaeus}, {Goldstein}, \&
  {Chhiber}}]{Usmanov2018}
{Usmanov}, A.~V., {Matthaeus}, W.~H., {Goldstein}, M.~L., \& {Chhiber}, R.
  2018, \href{http://dx.doi.org/10.3847/1538-4357/aad687}{\JournalTitle{\apj},
  865, 25}

\bibitem[{{van Ballegooijen} \&
  {Asgari-Targhi}(2016)}]{vanBallegooijenAsgariTarghi2016}
{van Ballegooijen}, A.~A., \& {Asgari-Targhi}, M. 2016,
  \href{http://dx.doi.org/10.3847/0004-637X/821/2/106}{\JournalTitle{\apj},
  821, 106}

\bibitem[{{van Ballegooijen} \&
  {Asgari-Targhi}(2017)}]{vanBallegooijenAsgariTarghi2017}
---. 2017,
  \href{http://dx.doi.org/10.3847/1538-4357/835/1/10}{\JournalTitle{\apj}, 835,
  10}

\bibitem[{{van der Holst} {et~al.}(2010){van der Holst}, {Manchester},
  {Frazin}, {V{\'a}squez}, {T{\'o}th}, \& {Gombosi}}]{vanderHolst2010}
{van der Holst}, B., {Manchester}, W.~B., I., {Frazin}, R.~A., {et~al.} 2010,
  \href{http://dx.doi.org/10.1088/0004-637X/725/1/1373}{\JournalTitle{The
  Astrophysical Journal}, 725, 1373}

\bibitem[{{van der Holst} {et~al.}(2014){van der Holst}, {Sokolov}, {Meng},
  {Jin}, {Manchester}, {T{\'o}th}, \& {Gombosi}}]{vanDerHolst2014}
{van der Holst}, B., {Sokolov}, I.~V., {Meng}, X., {et~al.} 2014,
  \href{http://dx.doi.org/10.1088/0004-637X/782/2/81}{\JournalTitle{\apj}, 782,
  81}

\bibitem[{{Verdini} {et~al.}(2019){Verdini}, {Grappin}, \&
  {Montagud-Camps}}]{Verdini2019}
{Verdini}, A., {Grappin}, R., \& {Montagud-Camps}, V. 2019,
  \href{http://dx.doi.org/10.1007/s11207-019-1458-y}{\JournalTitle{\solphys},
  294, 65}

\bibitem[{{Verdini} \& {Velli}(2007)}]{VerdiniVelli2007}
{Verdini}, A., \& {Velli}, M. 2007,
  \href{http://dx.doi.org/10.1086/510710}{\JournalTitle{\apj}, 662, 669}

\bibitem[{Vourlidas {et~al.}(2016)Vourlidas, Howard, Plunkett, Korendyke,
  Thernisien, Wang, Rich, Carter, Chua, Socker, Linton, Morrill, Lynch, Thurn,
  Van~Duyne, Hagood, Clifford, Grey, Velli, Liewer, Hall, DeJong, Mikic,
  Rochus, Mazy, Bothmer, \& Rodmann}]{Vourlidas2016}
Vourlidas, A., Howard, R.~A., Plunkett, S.~P., {et~al.} 2016,
  \href{http://dx.doi.org/10.1007/s11214-014-0114-y}{\JournalTitle{Space
  Science Reviews}, 204, 83}

\bibitem[{{Weber}(1970)}]{Weber1970}
{Weber}, E.~J. 1970,
  \href{http://dx.doi.org/10.1007/BF00963958}{\JournalTitle{\solphys}, 13, 240}

\bibitem[{{Weber} \& {Davis}(1970)}]{WeberDavis1970}
{Weber}, E.~J., \& {Davis}, L., J. 1970,
  \href{http://dx.doi.org/10.1029/JA075i013p02419}{\JournalTitle{\jgr}, 75,
  2419}

\bibitem[{{Weber} \& {Davis}(1967)}]{WeberDavis1967}
{Weber}, E.~J., \& {Davis}, Jr., L. 1967,
  \href{http://dx.doi.org/10.1086/149138}{\JournalTitle{\apj}, 148, 217}

\end{thebibliography}

\appendix
\section{Erratum}

\subsection{Equation of energy conservation}
\label{subsec:energy}

In the original version of the article, the source term of the total energy equation included an additional and unintended term $Q_w$. In fact, following the original notations, the conservation of the system's energy can be written equivalently in two ways:

\begin{align*}
\label{eq:En1}
\frac{\partial}{\partial t} (E + \mathcal{E} + \rho \Phi)  +& \nabla \cdot [(E+p+\rho \Phi)\mathbf{v}-\mathbf{B}(\mathbf{v}\cdot \mathbf{B}) + \mathbf{v}_g^+ \mathcal{E}^+ + \mathbf{v}_g^- \mathcal{E}^-] = Q - Q_w = Q_h-Q_c-Q_r, \numberthis
\end{align*}

or

\begin{align*}
\label{eq:En2}
\frac{\partial}{\partial t} (E + \rho \Phi)  +& \nabla \cdot [(E+p-p_w+\rho \Phi)\mathbf{v}-\mathbf{B}(\mathbf{v}\cdot \mathbf{B})] = Q - \mathbf{v} \cdot \nabla \frac{\mathcal{E}}{2}. \numberthis
\end{align*}
The form of these two equivalent equations can be understood as follows: when accounting for the conservation of both the wind energy and the waves (equation \ref{eq:En1}), the wave heating does not appear as a source but is instead hidden in the decay of the wave amplitude and energy. However, this term should appear when one only considers the fluid energy, as it is in equation \eqref{eq:En2}. Then, a term compensating for the wave pressure must be included. We chose to implement equation \ref{eq:En1}.

\subsection{New simulation of PSP encounter 1}
\label{sec:sim}

As a consequence of this redundant term, the wave heating was twice what it was meant to be. We consequently ran a new simulation using the correct energy equation. We chose to change slightly the input parameters to obtain a heating and wave amplitude very close to the original simulation. We increased the base velocity perturbation by 20\%, reaching the value:

\begin{equation}
    \delta v_{\odot} = 36 \; \mathrm{km/s},
\end{equation}
so that the total average input of Alfvén wave energy is $\langle \rho_{\odot} v_{A,\odot} \delta v_{\odot}^2 \rangle =  \rho_{\odot} \langle v_{A,\odot} \rangle \delta v_{\odot}^2 \approx 1.5 \times 10^{5}$ erg.cm$^{-2}$s$^{-1}$, with $\rho_{\odot} = 5 \times 10^{-16}$ g.cm$^{-3}$ and $\langle B_{\odot} \rangle \approx 1.8$G (the Alfvén wave flux at a given latitude and longitude depends on the precise value of the radial field).

We also decreased slightly the correlation length parameter to
\begin{equation}
    \lambda_{\odot} = 0.020 R_{\odot} \sqrt{\mathrm{G}} \approx 14 000 \; \mathrm{km}\sqrt{\mathrm{G}}.
\end{equation}

\begin{figure}
    \centering
    \includegraphics[width=7in]{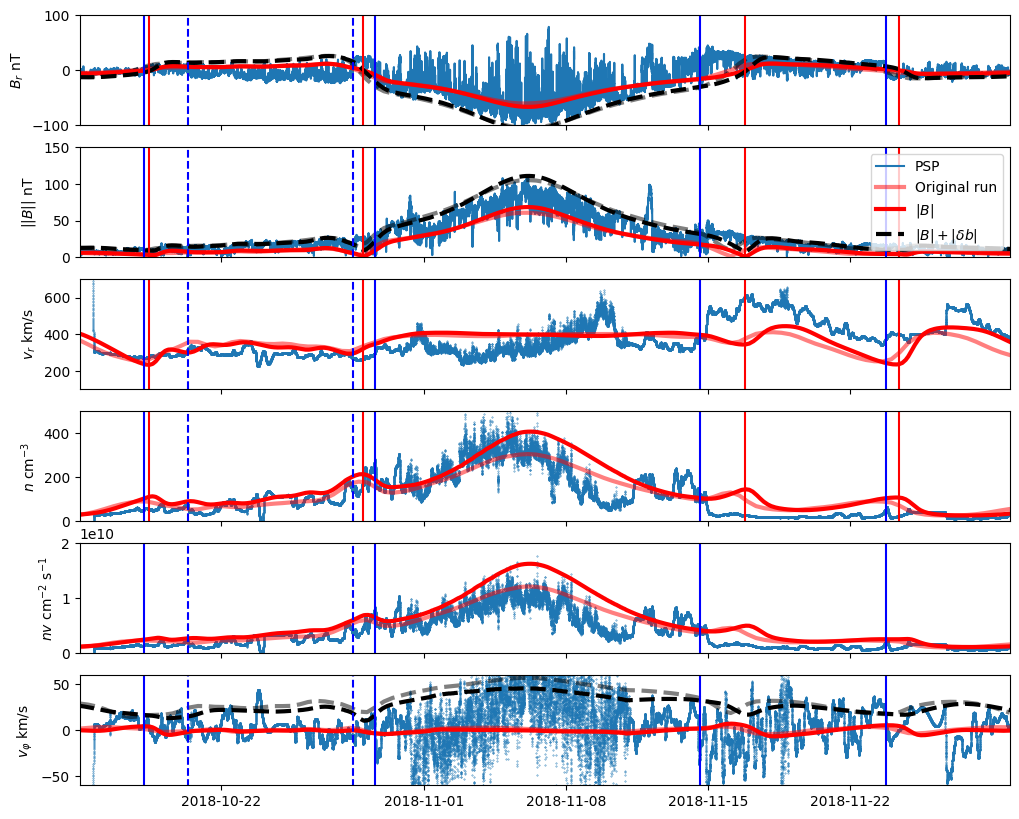}
    \caption{Comparison of the in situ measurements of PSP and the results of the new and original simulation. The original simulation results are displayed with transparency (both in red and dashed black). The vertical lines correspond to HCS crossings of the new simulation (in blue) and in the data (in red). The results of both simulations are very close, except for a slight increase in the solar wind density.}
    \label{fig:fig2_err}
\end{figure}

We now reproduce the figures that could have been modified using this new simulation. In Figure \ref{fig:fig2_err} (Figure 2 in the original paper), we reproduce the in situ observations of PSP E1 and compare with the results of the MHD simulations. We left the original run, playing with the transparency of the curve (alpha of $0.5$). We see that the in situ variables are only very slightly modified. The only notable difference is in the density, which can be up by 25\% at the perihelion compared to the original run. Both the original and the new run remain nonetheless compatible with the span in the observed density. The HCS crossings (vertical lines) remain similar, and the solar wind sources are thus not significantly modified. 
\begin{figure}
    \begin{tabular}{cc}
    \includegraphics[width=3.2in]{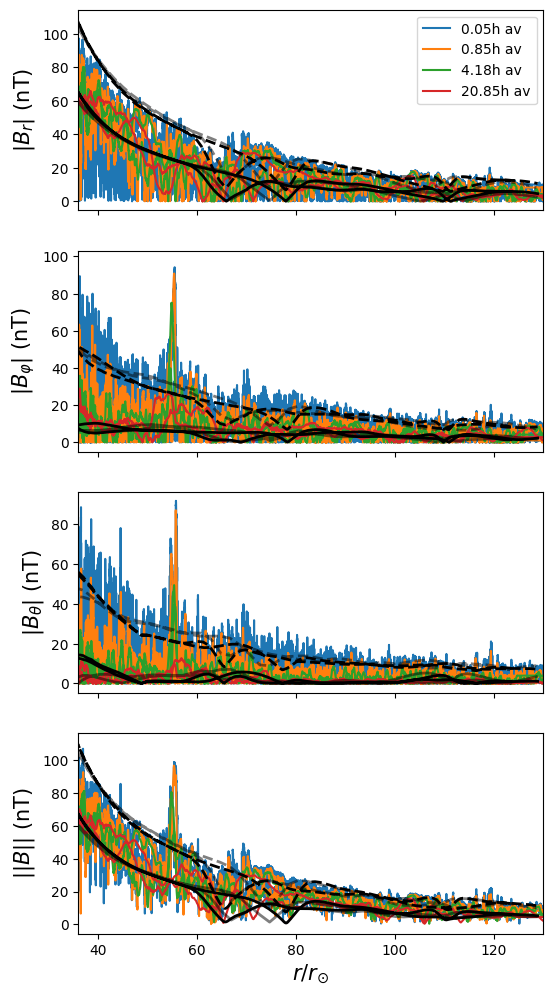} & \includegraphics[width=3.2in]{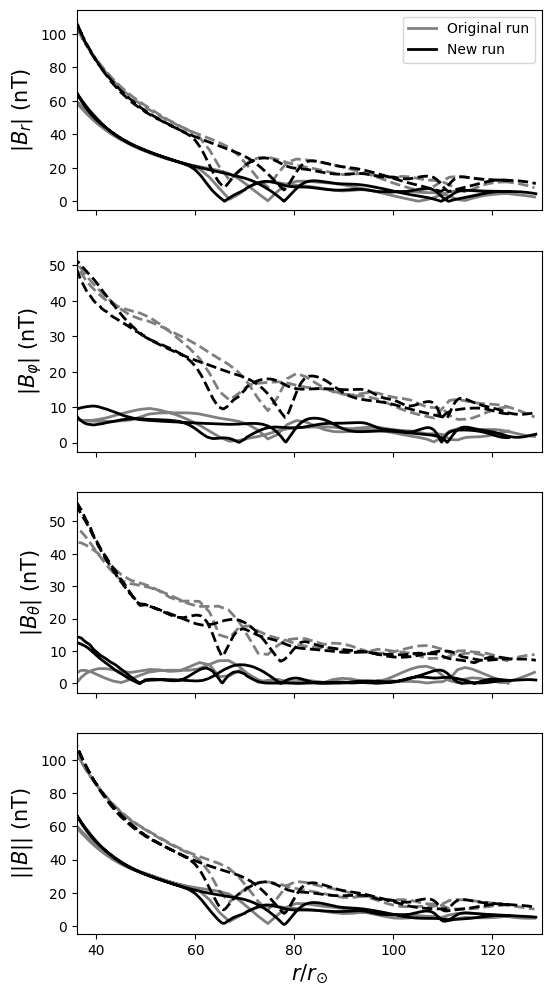} \\
    \end{tabular}
    \caption{Left panel: magnetic field measurements, with various running average timescales, as a function of the radial distance. The largest time average fits fairly well with the radial dependency of the field obtained in the simulation, shown in black. The dashed line illustrates the amplitude of the field when the Alfvén waves are accounted for. The original simulation results are displayed in transparency. In the right panel, we removed the data to better allow a comparison between the original and the new simulation.}
    \label{fig:fig4_err}
\end{figure}

In Figure \ref{fig:fig4_err} (Figure 4 in the original paper), we show the amplitude of the magnetic field as a function of distance in the data and the simulations. Again both simulation are very close, and as in the original paper, the amplitude of the radial magnetic field perturbations (switchbacks) fits with the amplitude of the waves in the simulation. 

Consequently, as shown with the novel simulation, the main conclusions of the original paper are unchanged:

\begin{itemize}
    \item Alfvén wave driven models of the solar corona can reproduce most in situ observables of the first Parker Solar Probe encounter of November 2018, to the notable exception of the tangential velocities.
    \item The amplitude of the perturbations necessary to power such a model are consistent with observations down to $35 R_{\odot}$.
    \item This includes perturbations in the radial magnetic field, i.e., switchbacks, that must then be a significant component of solar wind turbulence. 
\end{itemize}

Two following works have been impacted: \citet{Hazra2021} and \citet{Reville2020ApJL}. In both cases, a similar small parameter shift should yield results very close to the one published.

\end{document}